\documentclass[11pt,a4paper]{article}
\pdfoutput=1
\usepackage{jheppub}
\usepackage{amsfonts}
\usepackage{bbm}
\usepackage{graphicx}
\usepackage{caption}
\usepackage{subcaption}
\usepackage{stackrel}
\usepackage{mathtools}

\def\a{\alpha}
\def\b{\beta}
 
\def\d{\delta}

\def\i{\iota}

\def\l{\lambda}

\def\s{\sigma}                                  
\def\t{\tau}

\def\z{\zeta}
\def\vth{\vartheta}

\def\I{\Iota}

\def\inn{\,\in\,}
\def\AP#1{\mathcal A_{#1}}
\def\APhat#1{\wh{\mathcal{A}}_{#1}}

\def\ALhat{\widehat{\mathcal A}_{2,m}}

\def\ALstar{{\mathcal A}^\star_{2,m}}

\def\mypmod#1{\; (\text{mod} \; {#1})}
 \def\Av#1#2{\text{Av}^{(#1)}\bigl[#2\bigr]}
 \def\AvB#1#2{\text{Av}^{(#1)}\Bigl[#2\Bigr]}

\def\erfc{\text{erfc}}
\def\erf{\text{erf}}
\newcommand{\bal}{\begin{align}}
\newcommand{\eal}{\end{align}}

\newcommand{\cA}{\mathcal{A}}

\newcommand{\cC}{\mathcal{C}}

\newcommand{\cE}{\mathcal{E}}

\newcommand{\cH}{\mathcal{H}}
\newcommand{\cI}{\mathcal{I}}
\newcommand{\cJ}{\mathcal{J}}

\newcommand{\cL}{\mathcal{L}}
\newcommand{\cM}{\mathcal{M}}
\newcommand{\cN}{\mathcal{N}}

\newcommand{\cZ}{\mathcal{Z}}

\newcommand{\CC}{\mathcal{C}}

\newcommand{\CN}{\mathcal{N}}

\newcommand{\CZ}{\mathcal{Z}}

\newcommand{\CH}{\mathcal{H}}

\renewcommand{\Im}{{\rm Im}}
\renewcommand{\Re}{{\rm Re}}
\newcommand{\Tr}{\mbox{Tr}}

\newcommand{\sign}{\mbox{sign}}

\newcommand{\IR}{\mathbb{R}}
\newcommand{\IC}{\mathbb{C}}
\newcommand{\IZ}{\mathbb{Z}}
\newcommand{\IN}{\mathbb{N}}
\newcommand{\IH}{\mathbb{H}}

\newcommand{\half}{\frac{1}{2}}

\newcommand{\nn}{\nonumber}

\def\i{\mathrm{i}}

\def\p{\partial}

\def\bea{\begin{eqnarray}}
\def\eea{\end{eqnarray}}
\def\be{\begin{equation}}
\def\ee{\end{equation}}
\def\ba{\begin{align}}
\def\ea{\end{align}}

\newcommand{\bem}{\begin{pmatrix}}
\newcommand{\eem}{\end{pmatrix}}

\def\={\;  = \;}
\def\+{\, + \,}

\def\wt{\widetilde}
\def\wh{\widehat}
\def\bar{\overline}

\def\rt2{\sqrt{2}}

\renewcommand{\Im}{\mbox{Im}}
\renewcommand{\Re}{\mbox{Re}}

\def\I{\i}
\def\de{{\rm d}}
\def\tR{\widetilde R}

\newcommand{\ddh}{d}

\newcommand{\vv}{v} 
\newcommand{\MTN}{\cM_{\rm TN}}

\title{Mock modularity from black hole scattering states}

\author[a]{Sameer Murthy}
\author[b]{and Boris Pioline}
\affiliation[a]{Department of Mathematics, King's College London,\\
The Strand, London WC2R 2LS, UK.}
\affiliation[b]{\it Laboratoire de Physique Th\'eorique et Hautes Energies (LPTHE), \\
Sorbonne Universit\'e et CNRS UMR 7589, 4 place Jussieu, F-75005, Paris, France}
\abstract{The exact degeneracies of~quarter-BPS dyons in Type II string theory on $K3 \times T^2$ are 
given by Fourier coefficients of the inverse of the Igusa cusp form. For a fixed magnetic charge invariant~$m$, the 
generating function of these degeneracies naturally  decomposes as a sum of two parts, which are 
supposed to account for single-centered black holes, and two-centered black hole bound states, respectively. 
The decomposition is such that each part is separately modular covariant but neither is holomorphic,
calling for a physical interpretation of the non-holomorphy.
We resolve this puzzle by computing the supersymmetric index of the quantum mechanics of 
two-centered half-BPS black-holes, which we model by geodesic motion on Taub-NUT space 
subject to a certain potential. 
We compute a suitable index using localization methods, and find that it includes both
a temperature-independent contribution
from BPS bound states, as well as a temperature-dependent contribution 
due to a spectral asymmetry in the continuum of scattering states. The continuum contribution
agrees precisely with the non-holomorphic completion term required for the modularity of the 
generating function of two-centered black hole bound states.
}

\begin{document}

\maketitle

\section{Introduction and summary \label{sec:intro}}

The statistical explanation of thermodynamic entropy of black holes is one of the remarkable achievements
of string theory \cite{Sen:1995in,Strominger:1996sh}. The emerging picture is that a black hole is a
bound state of an ensemble of fluctuating strings, branes, and other fundamental excitations of 
string or M theory. This picture has been checked to great precision for supersymmetric black holes in 
superstring theory. The microscopic degeneracy in this case is captured by a  
supersymmetric index that counts all micro-states carrying the 
same charges as that of the black hole in a weakly coupled regime. 
This index is robust under small changes of moduli, which allows us to extrapolate the
weak coupling result to strong coupling. The leading order result for the logarithm of the
index at large charges is then found to match the thermodynamic black hole entropy, with no
adjustable parameter. 
This match can be pushed to higher order by computing and comparing the subleading 
corrections to both the macroscopic and microscopic results (see the review~\cite{Mandal:2010cj}). 

One can go even further and try to compute the exact macroscopic quantum entropy of  
supersymmetric black holes using the formulation of quantum entropy in \cite{Sen:2008vm}, 
and compare it with the logarithm of 
the microscopic degeneracy of states. For four-dimensional black holes  preserving four supercharges
in~$\CN=8$ string theory in asymptotic flat space (Type II string theory compactified on~$T^6$), 
one can actually sum up all the macroscopic quantum corrections using localization and recover 
the exact microscopic integer degeneracy \cite{Dabholkar:2011ec,Dabholkar:2014ema}. 
This result prompts us to look for such exact 
agreement in other systems and, in particular, in theories with less supersymmetry.

A crucial guide in this successful comparison of the exact microscopic and macroscopic entropy
is the modular symmetry of the 
generating function of the degeneracies of BPS states \cite{Maldacena:1997de}. 
The microscopic degeneracies\footnote{{In this paper the word 
`degeneracy' refers to a  suitable helicity supertrace 
that counts the net number of short multiplets with given charges. 
Under favorable circumstances, this may coincide with
the actual number of states~\cite{Sen:2009vz,Dabholkar:2010rm}.}} 
of~$\frac18$-BPS states 
in~$\CN=8$ string theory are Fourier coefficients of 
the ratio of powers of the Jacobi theta function and of the Dedekind eta 
function \cite{Maldacena:1999bp,Shih:2005qf,Shih:2005he,Pioline:2005vi}
$Z^{\CN=8}_\text{micro}(\t,z) = \vth_1(\t,z)^2/\eta(\t)^6$.
The function~$Z^{\CN=8}_\text{micro}$ is a weak Jacobi form \cite{MR781735} and, in particular, 
transforms covariantly under the modular group $SL_2(\IZ)$. This modular transformation property leads to an analytic 
formula for the microscopic degeneracy, known as the Hardy-Ramanujan-Rademacher expansion,
which expresses the integer coefficient of a modular form as an infinite series of Bessel functions of 
exponentially decreasing magnitude. This series can be interpreted on the macroscopic side
as an infinite sum over orbifold geometries with the same $AdS_2$ asymptotics \cite{Banerjee:2008ky,Murthy:2009dq}, 
and each term in the sum can be recovered, using localization, as the functional integral of bulk supergravity fluctuations 
around the corresponding saddle point~\cite{Dabholkar:2010uh,Dabholkar:2011ec,Dabholkar:2014ema}.

For the next-to-simplest case of~$\frac14$-BPS black holes in~$\CN=4$ 
string theories, it turns out, however, that the modular symmetry is not manifest.
The microscopic degeneracy is again a Fourier coefficient of a certain automorphic form, namely the 
inverse of the Igusa cusp form discussed below, but it includes contributions both from a single, 
spherically symmetric BPS black hole  as well as contributions from two-centered black hole 
bound states \cite{Sen:2007vb}.\footnote{{In~$\CN=8$ string vacua, multi-centered 
configurations have too many fermionic zero-modes to contribute  to the 
relevant spacetime helicity supertrace \cite{Dabholkar:2009dq}.}} 
In order to single out the single-centered black hole microstates, we need to remove part of the 
spectrum, thereby spoiling some of the symmetries. 
The observation of~\cite{Dabholkar:2012nd} was that the  modular 
symmetry is not broken, but has an anomaly:  the degeneracies of microstates of $\frac14$-BPS 
black holes are coefficients of~\emph{mock Jacobi forms}, which are holomorphic  but not modular. 
They can, however, be made modular at the cost of adding a 
correction term which is non-holomorphic in~$\t$ (but still holomorphic in~$z$)~\cite{Zwegers:2008zna}.  
This characterization allows one to generalize the Rademacher expansion and 
enables complete control over the growth of the Fourier coefficients~\cite{BringmannOno, BringmannMahlburg, Bringmann:2010sd}. 
It has also been used to make progress on the bulk interpretation of the microscopic 
degeneracies of black holes~\cite{Murthy:2015zzy, Ferrari:2017msn}. We note that a similar
phenomenon arises in the context of $\cN = 2$ black holes \cite{Manschot:2009ia,Alexandrov:2016tnf}, 
but in that context mock modular forms of higher depth are expected to arise 
due to the occurrence of BPS bound states
involving an arbitrary number of constituents \cite{Manschot:2010xp,AlexandrovBPtoappear}. 

In this paper, inspired by earlier work \cite{Alexandrov:2014wca,Pioline:2015wza} in the context
of $\cN = 2$ black holes, we attempt to give a physical justification of this non-holomorphic correction from 
the macroscopic point of view in the $\cN=4$  context, by computing the contribution of the continuum of 
scattering states in the quantum mechanics of two-centered BPS black holes. The rest of the introduction 
contains a summary of the details of our problem and its proposed solution.

\subsection{Dyon degeneracy function in~$\CN=4$ string theory and its decomposition}

Consider Type II string theory on~$K3 \times T^2$, a theory with four-dimensional~$\CN=4$ supersymmetry. 
The~$U$-duality group of the theory is~$SO(22,6,\IZ) \times SL(2,\IZ)$ \cite{Font:1990gx,Sen:1994fa}. 
There are~28 gauge fields with respect 
to which we have electric charges~$N^i$ and magnetic charges~$M^i$, $i=1,\cdots, 28$.
These charges transform as a vector under the T-duality group~$SO(22,6, \IZ)$, and the electric and magnetic 
charges transform as a doublet under the~$S$-duality group~$SL(2,\IZ)$. 
The T-duality invariants are~$(N^{2}/2, M \cdot N, M^2/2) \equiv (n, \ell, m)$, where the inner 
product is with respect to the~$SO(22,6, \IZ)$-invariant metric. 
The degeneracy of~$\frac14$-BPS dyons in this theory depends on these T-duality invariants as well as the 
point~$\phi$ in moduli space. 
The degeneracy is given as the Fourier coefficient \cite{Dijkgraaf:1996it,Shih:2005uc,David:2006yn}:
\be\label{dyondeg}   
d^\text{dyon}(n, \ell,m; \phi)
\= \int_{ \CC(\phi)} \, \frac{e^{-2\pi\I  (n \t + \ell z + m \s)}}{\Phi_{10}(\t, z, \s)} \, \de \t \, \de z \, \de \s \,,
\ee
where~$\Phi_{10}$ is the Igusa cusp form, the unique Siegel modular form of weight~$10$. 
Here the contour~$\CC$ depends on the moduli $\phi$ 
as well as the charge invariants (which we have 
suppressed in the above formula) \cite{Cheng:2007ch,Banerjee:2008yu} (see \cite{Bossard:2016zdx} 
for a recent new perspective on this formula).

Above we have used the terminology of ``dyon degeneracy" as is common, but it should be understood
that the left-hand side of the formula~\eqref{dyondeg} refers to the index of states that preserve a quarter of the 
spacetime supersymmetry. In the near-horizon region of attractor black holes, it turns out that all the states 
that contribute to this index are bosonic and therefore this index is really a degeneracy \cite{Sen:2009vz,Dabholkar:2010rm}, 
but more generally there can be cancellations between bosons and fermions. In particular one can show that the 
only gravitational configurations that have non-zero contributions 
to the supersymmetric index in this situation are 1) single-centered~$\frac14$-BPS dyonic black holes and 2) 
two-centered black holes, each of which is individually~$\frac12$-BPS \cite{Sen:2007vb}.  
This suggests that the generating function can itself be decomposed as a sum of single-centered black holes 
and two-centered black hole bound states. This intuition was made precise in the M-theory limit in~\cite{Dabholkar:2012nd}, 
in which we must first expand the generating function in the region~$\s \to \I\infty $\footnote{In the M-theory limit, 
following the contour~$\CC$ in~\eqref{dyondeg} leads to $\text{Im}(\t) = \frac{c_\t}{R}$, $\text{Im}(z) = \frac{c_z}{R}$, 
$\text{Im}(\s) = c_\s R$, with~$R \to \infty$, where~$c_\t$, $c_z$, $c_\s$ are functions of charges and other moduli that are held fixed in the limit, 
such that $\Im z=-\frac{\ell}{2m}\Im\t$.
\label{foo2}}:
\be\label{reciproigusa} 
\frac 1{\Phi_{10}(\s, \t, z)} \= \sum_{m=-1}^{\infty} \psi_m (\t,z) \, e^{2\pi\I  m \s}  \,.
\ee
The Fourier-Jacobi coefficients~$\psi_m (\t,z)$ are meromorphic Jacobi forms of weight~$-10$ and index~$m$ with a 
double pole at $z=0$ and no others (up to translation by the period lattice~$\IZ \t + \IZ$). 
The meromorphy is a hallmark of a phenomenon known as wall-crossing: as we vary $\Im(z)$ space,  
the Fourier coefficients of the Jacobi form~$\psi_m (\t,z)$ with respect to $\Re(z)$  jump when~$\text{Im}(z)$ crosses an integer 
multiple of~$\text{Im}(\t)$. This corresponds precisely to the appearance or disappearance of 
the bound state of two~$\frac12$-BPS black holes across a real codimension-one wall in the space
of moduli $\phi$, and the jump in the right-hand side of \eqref{dyondeg} is 
precisely the degeneracy carried by that bound state~\cite{Dabholkar:2007vk,Sen:2007vb}. 

Focussing on the case $m>0$ relevant for genuine black holes, 
the contributions  of two-centered bound states are captured by  
the  function~$\psi_{m}^{\rm P}(\t,z)$, called the \emph{polar part} of~$\psi_{m}$
constructed to have the same poles and residues as~$\psi_{m}$ as well as the same 
elliptic transformations under shifts of~$z$ by~$\IZ\tau+\IZ$. 
Its explicit form is given by
\be \label{psimP} 
\psi_{m}^{\rm P}(\t,z) \= \frac{\ddh(m)}{\eta(\t)^{24}} \; \AP{2,m}(\t,z) \, , 
\ee
where~$\ddh(m)$ is the 
$m^\text{th}$ Fourier coefficient of~$\frac{1}{\eta(\t)^{24}}$, which gives the degeneracy of half-BPS black holes 
with ${\rm gcd}(N^2/2,M\cdot N,M^2/2)=m$~\cite{Dabholkar:1989jt}, and 
$\AP{2,m}(\t,z)$ is the \emph{Appell-Lerch sum}
\be \label{defAL2} 
\AP{2,m}(\t,z)\= \sum_{s\in\IZ} \frac{q^{ms^2 +s} \, \z^{2ms+1}}{(1 -q^s \z)^2} \,, 
\qquad q = e^{2\pi\I  \t} \, , \quad \z = e^{2\pi\I  z} \, .
\ee
The contribution of single-centered black holes can be computed by evaluating~\eqref{dyondeg}
at the attractor point $\phi_*$. Following the contour $\cC(\phi_*)$ in the M-theory limit
(see Footnote \ref{foo2}), we are led to the generating function~$ \psi_{m}^{\rm F}(\tau,z)$, 
called the \emph{finite part} of~$\psi_{m}$, defined as:
\be
\psi_{m}^{\rm F}(\t,z) \= \sum_{\ell \; \text{mod}  2m} f_{m,\ell}^\text{attr}(\t) \, \vartheta_{m,\ell}(\t,z)  \, ,
\ee
where 
\be\label{fmlpsim}
f_{m,\ell}^\text{attr}(\t)  \=  e^{-\pi \I \ell^2 \t / 2m} \int_{- \ell \t/2m}^{- \ell \t/2m+1} \psi_m(\t,z) \, e^{- 2\pi\I  \ell z} \, dz \, , 
\ee 
and the weight-$\frac12$ index-$m$ theta function is defined as:
\begin{eqnarray} \label{thetadef} 
\vth_{m,\ell}(\t, z)  \= \sum_{{\l\in\IZ} \atop {\l\,\equiv\,\ell \mypmod{2m}}} q^{\l^2/4m} \, \z^\l \,.
\end{eqnarray}

With these definitions, one can now check that the meromorphic Jacobi form~$\psi_{m}(\t,z)$ is the sum 
of its finite and polar parts~\cite{Dabholkar:2012nd}:
\be \label{ThmDecomp} 
\psi_{m}(\tau,z) \=   \psi_{m}^{\rm F}(\tau,z) \, + \, \psi^{\rm P}_{m}(\tau,z) \, .
\ee
Since the function~$\psi_{m}^{\rm P}(\tau,z)$ has the same poles and residues as~$\psi_m$, 
the function~$\psi_{m}^{\rm F}(\tau,z)$ is holomorphic in~$z$, consistently with its interpretation 
as the generating function of single-centered black holes degeneracies, which cannot exhibit any wall-crossing phenomena.

\subsection{Mock Jacobi forms and the holomorphic anomaly}

The nontrivial part of the above decomposition theorem is of course its implication for modularity. 
The additive decomposition of~$\psi_{m}$ breaks modularity of the individual pieces and, in particular, $\psi_{m}^{\rm F}(\t,z)$  
is not a Jacobi form any more. 
The theorem states that by adding a specific non-holomorphic correction term that we will discuss in Section~\ref{sec:BHbnd}
(see Equation~\eqref{psimhat}), to~$\psi^{\rm F}_{m}(\t, z)$,  one can obtain a non-holomorphic 
completion~$\wh{\psi^{\rm F}_{m}}(\t, z)$  which \textit{is} modular 
and transforms as a Jacobi form of weight~$-10$ and index~$m$.  
As the left-hand side of~\eqref{ThmDecomp} is a Jacobi form, it is clear that  
subtracting the same non-holomorphic correction term from~$\psi_m^\text{P}(\t,z)$ also gives a function~$\wh{\psi_{m}^\text{P}}(\t, z)$ 
that transforms as a Jacobi form of the same weight and index. In other words:
\be \label{ThmDecomp2} 
\psi_{m}(\tau,z) \=   \wh{\psi_{m}^{\rm F}}(\tau,z) \, + \, \wh{\psi_{m}^{\rm P}}(\tau,z) \, ,
\ee
where both summands are non-holomorphic but modular.

The failure of holomorphy of  the completions~$\wh{\psi_{m}^{\rm F}}(\t, z)$ and~$\wh{\psi_{m}^{\rm P}}(\t, z)$ 
is captured by the following equation (with $\tau_2=\Im(\tau)$):
\be\label{holanom}
 \tau_2^{3/2} \; \frac{\partial} {\partial \bar{\tau}}   \, \wh{\psi_{m}^{\rm F}}(\tau,z) \= 
 - \tau_2^{3/2} \; \frac{\partial} {\partial \bar{\tau}}   \, \wh{\psi_{m}^{\rm P}}(\tau,z) \= 
 \sqrt{\frac{m}{8 \pi \I}} \; \frac{\ddh(m)}{ \eta(\tau)^{24}} \,  
 \sum_{\ell \mypmod{2m}}  {\overline{\vth_{m,\ell}(\tau,0)}} \, \vartheta_{m,\ell} (\tau,z) \, .
\ee
The fact that the completed partition function~$\wh{\psi_{m}^{\rm F}}$ transforms like a holomorphic Jacobi form 
suggests that it should be identified with the elliptic genus of the five-dimensional black string that descends to the black hole 
upon compactification on a circle.
It was speculated in~\cite{Dabholkar:2012nd} that the non-holomorphic dependence
on~$\t$ is caused by the non-compactness of the target space of the $SCFT_2$, similar to the 
phenomenon studied in \cite{Troost:2010ud,Eguchi:2010cb,Ashok:2011cy}. 
Unfortunately, a detailed implementation of this idea has remained elusive. 
In this paper we focus instead on the two-centered piece~$\wh{\psi_{m}^{\rm P}}$, and investigate
the physical origin of its non-holomorphic dependence.

\subsection{Moduli space of two-centered black holes and continuum contribution}

Consider the $(n,\ell)$th Fourier coefficient~$d^\text{P}(n,\ell,m)$ of the function~$\psi_m^\text{P}(\t,z)$ \eqref{psimP} 
with respect to the potentials~$(\Re(\t),\Re(z))$. This coefficient depends on the value of~$\Im(z)$ because of the meromorphy 
of~$\psi_m^\text{P}$. For a given value of~$\Im(z)$, determined by the values of the moduli
at spatial infinity, it is expected to compute the Witten index 
of the supersymmetric quantum mechanics of  two-centered BPS black holes with total charge invariants~$(n,\ell,m)$.
This interpretation has been checked very precisely:  
for fixed magnetic charge invariant~$m$, 
the walls of marginal stability of two-centered bound states
in the M-theory limit precisely correspond to the poles in~$z$ of the Appell-Lerch sum~\eqref{defAL2}. 
All these walls can be mapped, by S-duality, to the wall at~$\Im(z)=0$ across which a \emph{basic two-centered bound state} 
consisting of a purely electric $\frac12$-BPS black hole with charge invariant~$n$ 
and a purely magnetic $\frac12$-BPS black hole of charge invariant~$m$ is created. The generating function of BPS 
states of this basic two-centered bound state is given by
\be \label{basicBndSt}
\psi_m^\text{basic}(\t,z) \= \ddh(m) \, \frac{1}{\eta^{24}(\tau)} \, \frac{\z}{(1-\z)^{2}} 
\; =: \; \sum_{n,\ell} d_m^\text{basic}(n,\ell;u_2) \, q^n \, \z^\ell \,,
\ee
where we have explicitly shown the dependence, discussed above, of the Fourier coefficient on $\Im(z)$,
through the variable~$u_2\coloneqq \Im(z)/\t_2$. 
The invariant~$\ell$ corresponds to the field angular momentum in this bound state configuration, and  
the pole in~$z$ corresponds to the (dis)appearance of this bound state across a wall of marginal stablity. 

As we describe in Section~\ref{sec:BHbnd}, the full generating function~$\psi_m^\text{P}(\t,z)$ is given by 
the sum over all S-duality images of this basic two-centered function~$\psi_m^\text{basic}(\t,z)$, 
and for this reason it is enough to focus our attention on the latter.
Its Fourier coefficient is computed by the supersymmetric index 
\be \label{defdP}
d_m^\text{basic}(n,\ell; u_2) \= \Tr^\text{basic, bound}_{(n,\ell,m)} (-1)^F \,,
\ee
where the trace is taken over the bound state spectrum of the quantum mechanics 
describing the basic two-centered black hole configuration at the given value of~$u_2$. 
Since these bound states are normalizable and discrete, the trace reduces to a sum over the supersymmetric 
ground states. 
The idea that we pursue in this paper is that the completed polar part~$\wh{\psi_{m}^{\rm P}}(\tau,z)$ 
should similarly arise from a supersymmetric partition function
\be\label{defdPhat}
\wh{d^\text{basic}_m}(n,\ell;\b, u_2)  \= \Tr^\text{basic, all}_{(n,\ell,m)} (-1)^F e^{-\b \cH}
\ee
which includes contributions of the full spectrum in this same quantum mechanics. 
Here $\b$ is the inverse temperature~and $\cH$ is the quantum Hamiltonian of the two-centered configurations.
The contributions of the bound state spectrum is of course independent of~$\b$ and
equal to~\eqref{defdP}, since only supersymmetric ground states with $\cH=0$ contribute,
but now there can be an additional contribution from the 
continuum spectrum, since the densities of bosonic and fermionic states need not be equal. 
We define the corresponding generating function~$\wh{\psi_m^\text{basic}}(\t,z)$,
where $\beta$ is identified with $4\pi\tau_2$. 
Averaging as before over all the S-duality images, 
we should recover the completed function~$\wh{\psi_{m}^{\rm P}}(\tau,z)$ in~\eqref{ThmDecomp2}.

The quantum dynamics of the two-centered black hole bound state is not completely understood. 
In the context of black holes in~$\CN=2$ string vacua, it is well-described by the quiver quantum mechanics 
with 4 supercharges introduced in~\cite{Denef:2002ru}, or more simply by the supersymmetric quantum 
mechanics on $\IR^3$  which arises on its Coulomb branch~\cite{Denef:2002ru},\cite{Lee:2011ph},\cite{Pioline:2015wza}. 
In that case, $\frac12$-BPS bound states arise from supersymmetric vacua in the quantum mechanics on $\IR^3$ 
describing the relative motion, while the 4 fermionic zero-modes come the center-of-motion degrees of freedom. {Similarly, 
in the~$\CN=4$ context relevant for this paper, one would like to construct an analogue of the supersymmetric 
quantum mechanics on $\IR^3$ with 8 supercharges, such that $8$ of the 12 fermionic zero-modes carried 
by $\frac14$-BPS bound states arise from  the center-of-motion degrees of freedom, while the remaining 4 
correspond to the unbroken supersymmetries in the quantum mechanics describing the relative motion.}

{While such a model does not appear to be documented in the literature, we shall obtain it by 
reducing a supersymmetric sigma model with 8 supercharges  on Taub-NUT space,} which is known to describe  
dyonic bound states in weakly coupled supersymmetric gauge theories \cite{Bak:1999da,Bak:1999ip,Weinberg:2006rq}. 
One considers the dynamics of two~$\frac12$-BPS dyons of 
charge $(Q_1,P_1)$ and $(Q_2,P_2)$ on a sublocus of the Coulomb branch where the corresponding 
central charge vectors are parallel, so that there are no static forces between the two dyons.  Factoring 
out the center of motion, the dynamics  captured by geodesic motion on the {reduced} 
monopole moduli space. 
When $P_1,P_2$ are associated to two consecutive nodes on the Dynkin diagram
associated to the gauge group $G$, this moduli space 
turns out to be the Taub-NUT manifold $\MTN$, with metric
\be
\label{dsTN}
\de s^2 \= 
\frac{H}{2} (\de\vec r)^2 + \frac{1}{2H}\left( \de \psi +\vec A \cdot \de\vec r \right)^2 \  ,\qquad
H(r) \= \frac{1}{R} + \frac{1}{|\vec{r}|} \,.
\ee
Here, $\vec{r} \in \IR^{3}$ is the relative position of the two dyons, $\psi \in [0,4\pi]$ is the relative angle associated to large 
gauge transformations, and $\vec A$ is a connection along the circle fiber parametrized
by $\psi$ such
that  $\partial_i H = \epsilon_{ijk} \partial_j A_k$. 
The parameter $R$ controls the radius of the circle fiber at infinity, and is proportional 
to the square 
of the magnetic charges, while the momentum along the circle fiber is identified with 
the  Dirac-Schwinger-Zwanziger pairing $Q_1 P_2 - Q_2 P_1$.
Away from the 
locus  where the corresponding central charge vectors are parallel, the  dynamics  is still given by geodesic
motion on $\MTN$, but now subject to a potential proportional to the squared 
norm of the (tri-holomorphic) Killing vector $\partial_\psi$, with a coefficient that we 
denote by~$\l^2$.

We find that the function~$\wh{\psi_m^\text{basic}}(\t,z)$ is indeed encoded in this quantum mechanical system, 
but in a subtle manner. We need to introduce a third parameter~$\wt u_2$, which corresponds to a  
three-variable generalization~\cite{Zwegers:2008zna, Zwegers:2011} of the two-variable Appell-Lerch sum in~\eqref{defAL2}. 
Upon identifying this third parameter with the coupling constant~$\l$ introduced above as~$\wt u_2 = - \l R$, 
we find that the Fourier coefficients of the three-variable function are reproduced by a suitable index in the 
above quantum mechanical system, but only in the attractor chamber where $\sign(u_2)= -\sign(\ell)$.
In particular, this index, which we introduce in Section~\ref{susyQM}, and compute by localization 
methods in Section~\ref{sec:IndexLoc}, precisely reproduces the non-holomorphic completion term 
that that is required for modularity.

\medskip

The plan of this paper is as follows. In Section~\ref{sec:BHbnd} we discuss the microscopic 
partition function of the black hole bound states, and how it can be understood as a sum of 
S-duality images of the basic bound state partition function. We then discuss the appearance 
of the Appell-Lerch sums and their non-holomorphic modular completions, and introduce
a three-variable generalization. In Section~\ref{sec:moduli} we discuss the supersymmetric 
quantum mechanical system which we use to model the dynamics of the basic black hole 
bound state and discuss a set of refined indices which get only contributions from short multiplets.
In Section~\ref{sec:IndexLoc} we compute the refined index using localization, and discuss the 
relation of this result to the microscopic partition functions for the black hole bound states. In
 Section~\ref{sec_disc} we summarize and discuss some puzzles and open questions.
Appendix~\ref{sec_ham} contains a suggestive attempt to compute the spectral asymmetry 
directly by Hamiltonian methods, eschewing a full analysis of the quantum mechanical  model.

\section{Black hole bound states and  Appell-Lerch sums \label{sec:BHbnd}}

In this section we explain the physics and the mathematics of the two-centered black hole bound state partition 
function~$\psi_{m}^{\rm P}(\t,z)$. Then we present some Fourier expansions of the Appell-Lerch sums. 
Finally we discuss the mathematics of the non-holomorphic parts in some detail.

\subsection{Basic two-centered black hole bound state and its decay}

We first consider a system of two $\frac12$-BPS black holes where one center has
purely electric charge $(\vec N,0)$ and the other 
purely magnetic charge $(0,\vec M)$. The degeneracy of the internal states carried by the first
center is ~$\ddh(n)\equiv p_{24}(n+1)$, which is the Fourier coefficient
of the generating function~\cite{Dabholkar:1989jt}  
\be
\frac1{\eta(\t)^{24}} \= \sum_{n=-1}^\infty \, \ddh(n)\, q^n \,.
\ee
By S-duality, the degeneracy of the internal states carried by the  second center is~$\ddh(m)$.  
Depending on the values of the moduli at infinity, the quantum mechanics of the relative
degrees of freedom  has either no supersymmetric
ground states, or $|\ell|$ of them, where $\ell=\vec M \cdot \vec N $ is the Dirac-Schwinger-Zwanziger 
product of the charges of the constituents, 
transforming as a multiplet of spin $(\ell-1)/2$ under spatial $SO(3)$ rotations \cite{Denef:2000nb}.
The tensor product of the configurational and internal degrees gives $ |\ell| \,\ddh(n)\, \ddh(m)\,$
BPS bound states of total charge $(M,N)$.

We now consider a generating function of degeneracies with fixed magnetic charge invariant $m$ and arbitrary 
electric charge invariants  $n$ and~$\ell$, with chemical potentials~$\tau$ and~$z$, respectively. In the chamber 
where only bound states with $\ell>0$ are allowed, the contribution of the above bound states is then 
\be\label{basicmulti}
\ddh(m) \cdot \frac{1}{\eta^{24}(\tau)} \, \cdot \sum_{\ell > 0} \, \ell \, \z^{\ell} \,,
\ee
where $\z = e^{2\pi\I  z} $.
In contrast, in the chamber where only bound states with $\ell<0$ are allowed, the contribution of the  bound states  is
\be\label{basicmulti2}
 \ddh(m) \cdot \frac{1}{\eta^{24}(\tau)} \, \cdot \sum_{\ell < 0} \, (-\ell) \, \z^{\ell} \,.
\ee
The first and second factors are the internal degeneracies of the half-BPS magnetic and electric centers, respectively, 
as explained above. The third factor in~\eqref{basicmulti} and
\eqref{basicmulti2}, taking into
account configurational degrees of freedom, is the Fourier expansion of the 
meromorphic function
\bea\label{basicmulti2}
 \frac{\z}{(1-\z)^{2}} & \= & \begin{cases} \phantom{-}\sum_{\ell > 0} \ell \, \z^\ell &\text{if $|\z|<1$,} \\
                           - \sum_{\ell < 0} \ell \, \z^\ell &\text{if $|\z|>1$,}\end{cases}
\=  \frac12\, \sum_{\ell \in \IZ} \, (\sign(\text{Im}(z)) + \sign(\ell))\, \ell \, \z^{\ell} \, .
 \eea
The basic wall-crossing of the theory is clear from the above two equations: for a fixed 
value of~$\ell$, the 
degeneracy  jumps across the wall~$\text{Im}(z)=0$,
which is the image in complex~$z$-space of the wall in moduli space across which the two-centered 
bound state with the given value of~$\ell$ decays or is created.

\subsection{S-duality and the sum over all wall-crossings}

In~$\CN=4$ string theory, one can map all the codimension-one walls of marginal stability in moduli space~\cite{Sen:2007vb}. 
The possible decays of a dyonic state with charge vector~$(\vec{N},\vec{M})$ are related by S-duality to the basic 
decay $(\vec{N},\vec{M}) \to (\vec{N},0) + (0,\vec{M})$ discussed in the previous subsection. 
The image of this basic decay under a S-duality transformation 
$\gamma=\biggl( \begin{matrix} a & b \\ c & d \end{matrix} \biggr)\in SL_2(\IZ)$ is 
\be \label{gendecay}
(\vec{N},\vec{M}) \to (ad \vec{N} - ab \vec{M}, cd \vec{N} - cb \vec{M}) + (-bc \vec{N} + ab \vec{M}, -cd \vec{N} +ad \vec{M}) \,,
\ee
where the two constituents have collinear electric and magnetic charges, as appropriate for $\frac12$-BPS states. 
The matrix $\gamma$ simultaneously acts linearly on the chemical potentials $(\rho,v,\sigma)$, 
mapping the wall $\Im(z)=0$ to $\Im(cd \tau+ (ad+bc) z + ab \sigma)=0$.

These walls can be mapped to the plane of the four-dimensional 
complex modulus\footnote{In terms of the string compactification, this is 
given by~$S=a+e^{-2\phi}$, where~$a$ and~$\phi$ are the heterotic axion and dilaton, respectively.}
$S=S_1+iS_2 \in \IH$. 
In the upper-half~$S$-plane, the walls are either straight lines intersecting the~$S_1$-axis
at the integers, or minor circular arcs intersecting the~$S_1$-axis at consecutive integers. 
The analysis of~\cite{Dabholkar:2012nd} is performed in the M-theory limit, in which the radius~$R$ of the M-theory 
circle is taken to be large keeping other scales in the problem fixed. In this limit, the modulus scales as~$S_2 \sim R$, 
and as a consequence, the only relevant walls in this limit are the straight lines.
The basic wall at~$z=0$ maps to~the vertical line at $S_{1}=0$. The other straight lines
are images of this line under the S-duality transformation $\gamma=\biggl( \begin{matrix} 1 & s \\ 0 & 1 \end{matrix} \biggr)$, 
$ s \in \IZ$, and are therefore associated to the decay
\be \label{Mthydecay}
(\vec{N},\vec{M}) \to (\vec{N} -  s \vec{M}, 0) + (s\vec{M},  \vec{M})  \qquad \text{at~$\text{Im}(z)= s\,\Im(\t)$} \,.
\ee
The number of configurational BPS ground states on a suitable side of this wall
is~$\vec{N_1} \cdot \vec{M_2} - \vec{N_2} \cdot \vec{M_1} = \ell - 2m s$, while
the electric charge invariant for the purely electric constituent is~$\vec{N_1}^2/2 = n + s^2m-s \ell$.
The S-duality transformation parameterized by the integer~$s$ can thus be identified with 
the elliptic transformation~$z \to z + s \t$ acting on Jacobi forms of index $m$.

The full generating function that captures all bound states relevant in the M-theory limit
is therefore obtained by summing over the elliptic transformation images of~\eqref{basicmulti2}. 
This is achieved by the operator:
\be \label{defAvm}  
\Av m{F(y)} \;\coloneqq\; \sum_{s \in \IZ} \,q^{ms^2} \, \z^{2ms} \, F(q^s \z)  \,
\ee
which sends any function of $\z$ of polynomial growth in~$\z$ 
to a function of $\z$ transforming like an index~$m$ Jacobi form under translations by the full 
lattice $\IZ\t+\IZ$~\cite{Dabholkar:2012nd}.  
Applying this to the function~\eqref{basicmulti2} leads to the Appell-Lerch sum:
\be \label{defBP} 
\AP{2,m}(\t,z)\=  \AvB m{ \frac{\z}{(\z-1)^2} } \= \sum_{s\in\IZ} \frac{q^{ms^2 +s} \, \z^{2ms+1}}{(1 -q^s \z)^2} \, .
\ee
The moduli dependence of the Fourier coefficients of Appell-Lerch sum $\AP{2,m}$ 
is apparent in the  
following Fourier expansion, valid when~$u_2 \equiv \Im(z)/\Im(\tau)$ is not an integer:
\be \label{BBm2}
\AP{2,m}(\t,z) \=  \frac12 \sum_{s\in\IZ} \sum_{\ell\in\IZ}
\, \Bigl( \, \sign ( s+u_2) + \sign(\ell) \, \Bigr) \,  \ell 
\, q^{ms^2 + \ell s} \, \z^{2ms+\ell} \, .
\ee
Note that the ambiguity of $\sign(\ell)$ at $\ell=0$ is irrelevant since this term does not contribute to the sum. 

Thus the final answer for the full generating function of two-centered black hole bound state degeneracies is precisely 
the polar part of meromorphic Jacobi form~$\psi_{m}$ discussed in the introduction:
\be \label{psimPavg} 
\psi_{m}^{\rm P}(\t,z) \= \frac{p_{24}(m+1)}{\eta(\t)^{24}} \; \AP{2,m}(\t,z) \, .
\ee

\subsection{Non-holomorphic modular completion}

The completion~$\APhat{2,m}$ of~$\AP{2,m}$ is defined as: 
\be 
\label{defGsmhat} 
\ALhat(\t,z) \= \AP{2,m}(\t,z)\;+\;\ALstar(\t,z) \,,
\ee
where
\be
\ALstar(\t,z) \; \coloneqq \; m\,\sum_{\text{\rm $\ell$ (mod $2m$)}}
\vth^{*}_{m,\ell}(\t)\,\vth_{m,\ell}(\t,z) \,,
\ee
with~$\vth^{*}_{m,\ell}$ given by the non-holomorphic Eichler integral of~$\vth_{m,\ell}$:
\be
\label{defEichler1/2} 
\vth^{*}_{m,\ell}(\t) \; \coloneqq \; \frac{\overline{\vth_{m,\ell}(\t)}}{2\pi\sqrt{m\t_2}} \,-\, 
\sum_{\l\inn\IZ+\ell/2m} |\l|\,\erfc\bigl(2|\l|\sqrt{\pi m\t_2}\bigr)\,q^{-m\l^2}  \,.
\ee
The completion~$\APhat{2,m}$ transforms as a Jacobi form of weight~2 and index~$m$~\cite{Zwegers:2008zna},\cite{Dabholkar:2012nd}. 
Given that~$1/\eta(\t)^{24}$ is a modular form of weight~$-12$, we have that completion of the two-centered 
generating function
\be \label{psimhat} 
\wh{\psi_{m}^{\rm P}}(\t,z) \; \coloneqq \; \frac{\ddh(m)}{\eta(\t)^{24}} \; \ALhat(\t,z) \, .
\ee

Putting together the above defining equations of~$\ALstar(\t,z)$, we can rewrite it as:
\be
\label{ALstar1}
\begin{split}
  \ALstar(\t,z) \= & 
\, m\,\sum_{\text{\rm $\ell$ (mod $2m$)}}\,
     \sum_{{r\in\IZ} \atop {r\,\equiv\,\ell \mypmod{2m}}} q^{r^2/4m} \, \z^r 
     \\
     & \; \times 
  \sum_{\l\inn\IZ+\ell/2m}
  \left[ \frac{1}{2\pi \sqrt{m\tau_2}} \bar q^{m \l^2} -
   |\l|\,\erfc\bigl(2|\l|\sqrt{\pi m\t_2}\bigr)\,q^{-m\l^2} \right] \, .
\end{split}
\ee
In this summation, $\ell'=2m\lambda$ runs over all integers, while 
the constraint $r\,\equiv\,\ell \mypmod{2m}$ is equivalent to $r\,\equiv\,\ell' \mypmod{2m}$.
We solve this constraint by setting $r=2ms+\ell'$ with $s\in\IZ$. Dropping the prime on $\ell'$,
we obtain
\be \label{ALstar2}
 \ALstar(\t,z) \= 
\sum_{s,\ell\in\IZ}
 \left[ \frac{1}{2\pi}\sqrt{\frac{m}{\tau_2}} \, e^{-\pi \ell^2 \tau_2/m} -
  \frac{ |\ell |}{2}\,\erfc\Bigl(|\ell|\sqrt{\frac{\pi\t_2}{m}}\Bigr) \right]\, q^{ms^2+\ell s}\, \z^{2ms+\ell} \,.
\ee
Combining Equations~\eqref{BBm2} and~\eqref{ALstar2},  the full completed
Appell-Lerch sum is given by
\be \label{ALhat}
\ALhat(\t,z) \= 
 \sum_{s,\ell\in\IZ}
 \left[\frac{1}{2\pi}\sqrt{\frac{m}{\tau_2}}  \, e^{-\pi \ell^2 \tau_2/m} +
  \frac{\ell }{2} \left( \erf\Bigl( \ell \sqrt{\frac{\pi\t_2}{m}}\Bigr) + \sign(u_2+s) \right)  \right]\, q^{ms^2+\ell s}\, \z^{2ms+\ell} \,.
\ee
In this form, the modular invariance of \eqref{ALhat} is a straightforward consequence of Vign\'eras's
criterion for the modularity of indefinite theta series~\cite{Vigneras:1977}.

\subsection{Three-variable Appell-Lerch sum \label{sec_AL3}}
The two-variable completed Appell-Lerch sum~\eqref{ALhat} can in fact be obtained 
by acting with a suitable derivative operator on the  weight-one
indefinite theta series with  two elliptic parameters 
\be
\label{ALhat13var}
 \wh{\cA}_{1,m}(\t,z,\wt z) \; \coloneqq \; \frac12  \sum_{s\in\IZ} \sum_{\ell\in\IZ}  \left[  \sign(s+u_2) + \erf
 \biggl( (\ell+\wt u_2) \sqrt{\frac{\pi\tau_2}{m}} \, \biggr) \right]  \, q^{ms^2+ \ell  s}\, \z^{2ms+\ell} \, \wt\zeta^s \,,
\ee
where $\wt\zeta=e^{2\pi\I  \wt z}$ and $\wt u_2=\Im(\wt z)/\tau_2$. The theta series
\eqref{ALhat13var} is closely related to Zwegers's Appell-Lerch sum $\mu(\tau,u,v)$,
see e.g.~\cite[\S 3.3]{Gupta:2018krl}. Indeed, a simple computation shows 
that
\bea
 \label{Aprm}
\wh{\cA}'_{1,m} (\t,z,\wt z)  &\; \coloneqq \; & \frac{1}{2 \pi \I} \bigl(\p_z - 2m \, \p_{\wt z} \bigr) 
\wh{\cA}_{1,m}(\t,z,\wt z) \nn  \\
&\=& \sum_{s\in\IZ} \sum_{\ell\in\IZ}\,\wh{a}_{\ell} (\t_2,u_2+s , \wt u_2) \,  q^{ms^2+\ell s}\, \z^{2ms+\ell} \, \wt\zeta^s
\eea
where
\be \label{aprmFour}
\wh{a}_{\ell} (\t_2, u_2, \wt u_2)  \; \coloneqq \;   \frac12 \,  \ell \biggl( \sign(u_2) + \erf
 \biggl( (\ell+\wt u_2) \sqrt{\frac{\pi\tau_2}{m}} \, \biggr) \biggr) 
 + \frac{1}{2\pi} \sqrt{\frac{m}{\t_2}} \; e^{-\pi\tau_2(\ell+\wt u_2)^2/m}   \, .
\ee
Thus, $\wh{\cA}'_{1,m} (\t,z,\wt z)$ reduces to  $\ALhat(\t,z)$ at $\wt z=0$. 

The quantity $\wh{a}_{\ell} (\t_2, u_2, \wt u_2)$ defined in \eqref{aprmFour}, 
which appears as the  Fourier coefficient of the term in \eqref{Aprm} with $s=0$, is
the one which we shall be able to obtain from an index computation in the 
supersymmetric quantum mechanics of the basic black hole bound state. More precisely, we 
shall identify its  value 
\be \label{amattr}
\wh{a}_{\ell}^\text{attr}(\tau_2,\wt u_2) \; \coloneqq \;  \frac12 \, \ell \, \biggl( -\sign(\ell) + \erf
 \biggl( (\ell+\wt u_2) \sqrt{\frac{\pi\tau_2}{m}} \, \biggr) \biggr) 
 + \frac{1}{2\pi} \sqrt{\frac{m}{\t_2}} \; e^{-\pi\tau_2(\ell+\wt u_2)^2/m}    \, .
\ee
at the attractor point~$u_2=-\ell/2m$ with a suitable index \eqref{Iplusqresult} 
receiving contributions both from discrete states and from the continuum of scattering states. 

We do not know yet how to recover~\eqref{aprmFour} away from the attractor chamber, 
since we have not been able to identify the effect of the variable $u_2=\Im z/\tau_2$
on the supersymmetric quantum mechanics. We note, however, that the 
Fourier coefficient in the non-holomorphic correction term~\eqref{ALstar2} is independent of $u_2$, 
and is entirely reproduced by the limit of~\eqref{amattr} as~$\wt u_2\to 0$, 
\be \label{amattrzero}
\wh{a}_{\ell}^\text{attr}(\tau_2,\wt u_2=0) \=  - \frac{ |\ell |}{2}\,\erfc\Bigl(|\ell|\sqrt{\frac{\pi\t_2}{m}}\Bigr)  
 + \frac{1}{2\pi} \sqrt{\frac{m}{\t_2}} \; e^{-\pi\tau_2 \, \ell^2/m}   \, .
\ee
Moreover, the Fourier coefficient the holomorphic 
two-variable Appell-Lerch sum~\eqref{BBm2} agrees with the limit 
\be\label{amattrinfty}
\lim_{|\wt u_2| \to \infty} \wh{a}_{\ell}^\text{attr}(\tau_2,\wt u_2) \=    \ell \, \bigl( \sign(\ell) - \sign(\wt u_2)  \bigr)  \, ,
\ee
upon formally identifying $\wt u_2$ with~$-u_2$. Thus, at a mathematical level
the full completed Appell-Lerch sum~\eqref{ALhat} can be recovered from the
quantum mechanics computation.

\section{Moduli space dynamics of two-centered black holes \label{sec:moduli}}

In this section we review the supersymmetric quantum mechanics that captures the relative low-energy 
dynamics of the dyonic bound states. The bosonic part corresponds to geodesic motion on Taub-NUT space, 
subject to a suitable potential. We briefly review the known spectrum of 
BPS bound states and the relevant indices which are sensitive to them.

\subsection{Classical dynamics of mutually non-local dyons}

As mentioned in the introduction, the relevant properties of $\frac14$-BPS black hole bound states 
in $\cN=4$ string vacua are  captured by the supersymmetric quantum mechanics describing 
the dynamics of two~$\frac12$-BPS dyons  in weakly coupled four-dimensional~$\CN=4$ Super Yang-Mills 
theories with gauge group $SU(3)$, carrying  magnetic charges associated to the two simple roots of $SU(3)$.
This problem has been intensively studied in the literature ~\cite{Bak:1999da, Bak:1999ip, Gauntlett:1999vc,Weinberg:2006rq} 
using a two-step procedure: first by considering a point on the Coulomb where the six adjoint Higgs fields in the Cartan 
algebra of $SU(3)$ are aligned, and then perturbing away from this locus. When the Higgs fields are aligned, 
the classical theory reduces to $SU(3)$ Yang-Mills theory with a single adjoint Higgs field. In this case, the 
two dyons do not experience any static forces, and their relative motion of two dyons with  is governed by 
{geodesic motion on  Taub-NUT space $\MTN$ with metric  \eqref{dsTN}. 
In units where  the reduced mass is set to 1, the Lagrangian is simply
\be
\label{LagGH}
\cL^\text{TN} \= 
\frac{H}{2} \left(\frac{\de\vec r}{\de t}\right)^2 + \frac{1}{2H}\left( \frac{\de \psi}{\de t} +\vec A \cdot \frac{\de\vec r}{\de t} \right)^2 \ .
\ee
where $H(r) \= \frac{1}{R} + \frac{1}{|\vec{r}|}$ and 
$\psi \in [0,4\pi]$ parametrizes the circle fiber at infinity.
}
Denoting by $\vec p$ and\footnote{The momentum $q$ in this section should not be confused
with the modular parameter $q=e^{2\pi\I\tau}$ in the previous section.} 
$q\in \IZ/2$ the canonical momenta conjugate to $\vec r$ and $\psi$, 
the Hamiltonian describing this geodesic motion is then 
\be
\label{hamTN}
\cH^\text{TN} \= \frac1{2H} \, (\vec p - q \vec A)^2 + \frac{H}{2} \, q^2\ ,
\ee
where $\vec A$ is the potential for a unit-charge Dirac monopole sitting at $\vec r=0$. The momentum $q$ is equal to half the Dirac-Schwinger-Zwanziger pairing of the two dyons, and we shall restrict 
our attention to $q\neq 0$, corresponding to the mutually non-local case. The potential $V
=\frac12 H q^2$ 
being monotonically decreasing towards spatial infinity, this system admits no bound states, but only scattering states.

Upon perturbing away from the single-Higgs field locus, it has been shown that the two dyons start 
experiencing static forces, such that their relative motion is described by motion on the same Taub-NUT 
space with an additional potential term proportional to the square of the Killing vector $\partial_\psi$. 
This potential being invariant under translations along the fiber, the momentum $q$ is still conserved 
and the relative dynamics is now described by the Hamiltonian
\be
\label{hamTNdef}
\cH \= \frac1{2H} \, (\vec p - q \vec A)^2 + \frac{H}{2} \, q^2+ \frac{\lambda^2}{2H} \,,
\ee
where $\lambda$ measures the distance away from the  single-Higgs field locus. At the classical level,
it is straightforward to see that the potential $V=\frac{H}{2} \, q^2+ \frac{\lambda^2}{2H}$ admits bound states 
whenever $|\lambda|>|q/R|$ is large enough, localized around the global minimum at
 \be
 \label{rcases}
 r_0 \= \begin{cases} \frac{q}{\lambda-q/R}, & \mbox{if}\ \lambda>q/R>0 \ \mbox{or}\  \lambda<q/R<0 \,, \\
 -\frac{q}{\lambda+q/R} & \mbox{if}\ \lambda> -q/R>0 \ \mbox{or}\  \lambda<-q/R<0 \,.
 \end{cases}
 \ee
 In either case, the ground state energy is $V(r_0)=|\lambda q|$ (independently of $R$), 
corresponding to a binding energy 
 \be
 \label{binding}
 \Delta E_0 = E_c - V(r_0) = \frac{R}{2} \vartheta_\pm ^2  \ ,\quad 
 \vartheta_\pm = \lambda \mp \frac{q}{R}\ ,
 \ee
 where $E_c=\lim_{r\to\infty} V(r) = \frac12  \bigl(\frac{q^2}{R}+\lambda^2 R \bigr)$. Note
 that \eqref{binding} holds provided  that
 bound states exist, namely $q \vartheta_+>0$ or $q \vartheta_-<0$, and 
 that the sign $\pm$ is equated with the sign of $q\lambda$.
 In addition,
 as in the case of the hydrogen atom, we expect an {infinite number of discrete}
 bound states with energy
 ranging between $E=|\lambda q|$ and $E_c$. If instead 
 $|\lambda|<|q/R|$ is too small, the potential is monotonically decreasing towards infinity, and 
there are no classical bound states. Thus, as the parameter $\lambda$ is varied from $-\infty$ to $+\infty$, 
bound states disappear when $\lambda$ crosses the value $-|q/R|$ and reappear when 
it crosses $|q/R|$. In addition, irrespective of the value of $\lambda$, the classical 
spectrum admits a continuum of scattering states with energy~$E\geq E_c$.  

\begin{figure}
\vspace*{-2cm}
\centerline{\includegraphics[width=0.5\textwidth]{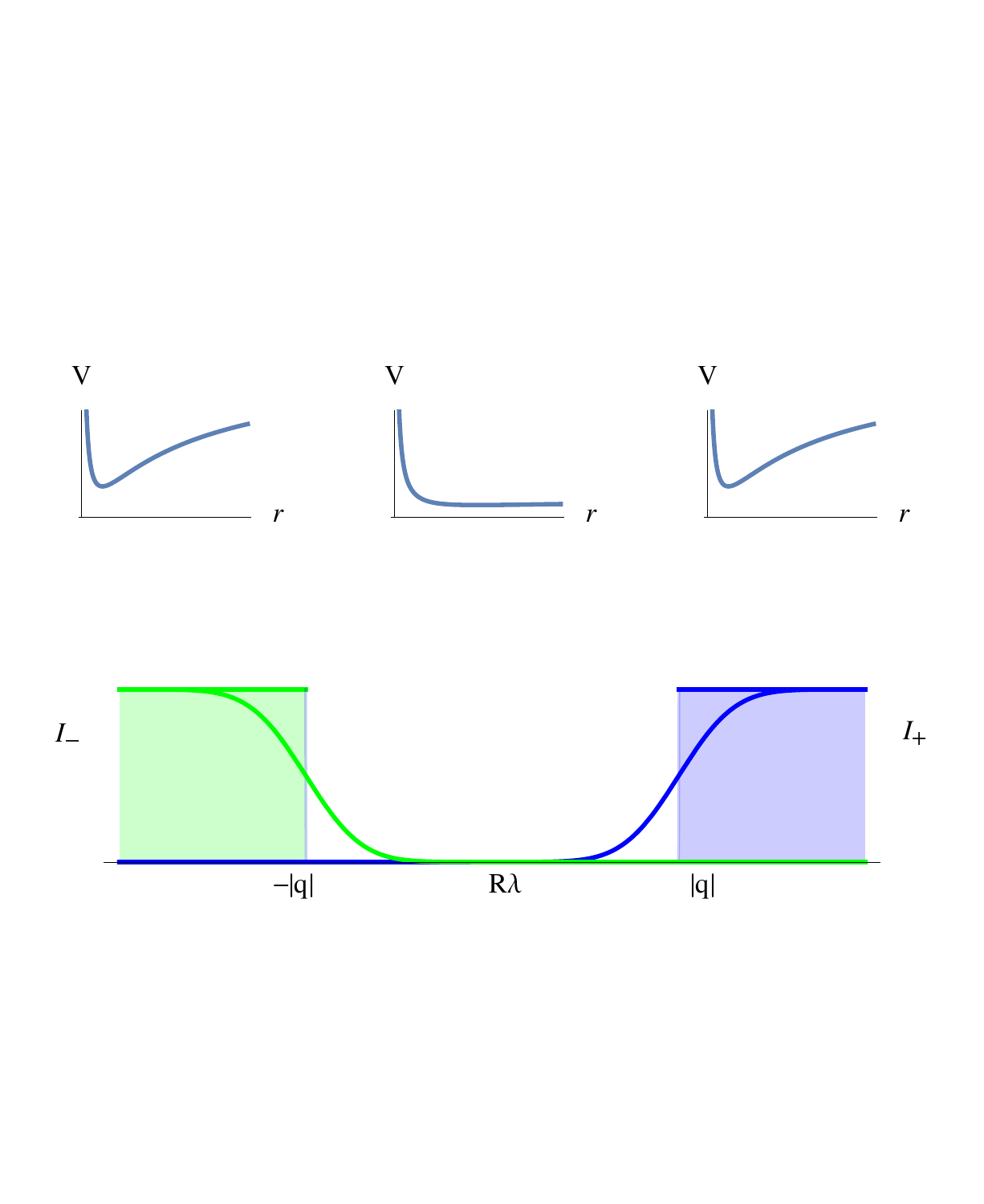}}
\vspace*{-2cm}
\caption{Top: Radial potential $V$ for $R\lambda<-|q|$, $-|q|<R\lambda<|q|$, and $R\lambda>|q|$, respectively. BPS bound 
states exist only for $R|\lambda|>q$. Bottom: indices $\cI^{\pm}$ and $\wh\cI^{\pm}(\beta)$ as
functions of $R\lambda$, for a fixed value of the inverse temperature.\label{fig}} 
\end{figure}

\subsection{Bosonic quantum mechanics}
We now briefly discuss the spectrum of the quantum Hamiltonian obtained by replacing $\vec p$
by $\I \partial/\partial \vec r$ in  \eqref{hamTNdef}. The resulting operator commutes with the angular momentum operator 
\be
\label{Jmag}
\vec J = \vec r \wedge  (\vec p - q \vec A)- q \frac{\vec r}{r} \ .
\ee 
In a sector with
$\vec J^2 = j(j+1)$ and $J_3=m$, the wave function
$\Psi(\vec r)$ factorizes into a radial part $f(r)$ and a monopole harmonic $Y_{q,j,m}$ with 
\be
\label{reljq}
j= |q|+\ell \,, \qquad  -j\leq m \leq j \,, \qquad m-q\in\IZ \,, 
\ee
with $\ell\in\IN$ the orbital angular momentum. 
The radial part of the Schr\"odinger
equation $\cH\Psi=E\Psi$ is then 
\be
\label{schrod}
\left[ \frac{1}{2H} \left( - \frac{1}{r} \partial_r^2 r + \frac{j(j+1)-q^2}{r^2} \right) 
+  \frac{H}{2} q^2 + \frac{\lambda^2}{2H} - E \right] \, f(r) \= 0 \,.
\ee
We shall denote $E=\frac{R}{2}k^2$ with $k\geq 0$
and $\vartheta=\sqrt{(q/R)^2+\lambda^2}$ such that the continuum starts at 
$E_c=\frac{R}{2}\vartheta^2$. Setting $f(r)=W(2\I  r \sqrt{k^2-\vartheta^2})$,
we find that \eqref{schrod} reduces to the Whittaker equation
\be
\left[ \partial_z^2 -\frac14 - \frac{R^2 k^2-2q^2}{2R\, z \sqrt{\vartheta^2-k^2}}
-\frac{j(j+1)}{z^2} \right]\, W(z) \= 0\ .
\ee  
The solutions are  linear combinations of Whittaker functions,
\be
\label{solwhit}
r\, f(r) \= \beta\, M_{\mu, \nu}\left( 2 r \sqrt{\vartheta^2-k^2}\right) + \gamma\, 
W_{\mu, \nu}\left( 2 r \sqrt{\vartheta^2-k^2}\right)
\ee
with
\be
\label{munu}
\mu\=-\frac{R^2k^2-2q^2}{2R\sqrt{\vartheta^2-k^2}}\ ,\qquad
\nu\=j+\frac12\ .
\ee
In order for the wave function to be regular at the origin, the coefficient $\gamma$ must vanish.
For normalizable bound states, the parameter $\mu$ (hence the radial wave number $k$)
can only take discrete values in order for the wave function to  decay at infinity. Using the standard formula
\be
\label{WtoM}
M_{\mu,\nu}(z) \= \frac{\Gamma(2\nu+1)}{\Gamma(\nu-\mu+\tfrac12)}\,
e^{\I\pi\mu}\, W_{-\mu,\nu}(e^{\I\pi}z)+
\frac{\Gamma(2\nu+1)}{\Gamma(\nu+\mu+\tfrac12)}
e^{\I\pi(\mu-\nu-\frac12)}\, W_{\mu,\nu}(z) \, ,
\ee
and $W(z)\sim z^\lambda e^{-z/2}$ as $|z|\to\infty$,  
we see that this happens when  $\Gamma(\mu+\nu+\frac12)$ has a pole, 
i.e.\footnote{This result for the discrete spectrum is in agreement with 
\cite[(E.30)]{Brennan:2018ura} upon identifying $(R,q,\lambda)$ in our notations
with $(R,p/2,C/\ell)$ in their notation, and fixing the scale $\mu=1/\ell$.}
\be
\frac{R^2 k_n^2-2q^2}{2R\sqrt{\vartheta^2-k_n^2}} \= j+ n +1 \ ,\quad n\in \mathbb{N}\ ,
\ee
where we recall that $j=|q|+\ell$. As expected in a bosonic model, the ground state $\ell=n=0$,
transforming as a spin $|q|$ representation of $SU(2)$, have
energy strictly bigger than the minimum $V(r_0)=|q\lambda|$ of the potential .

In contrast, for  scattering states, the radial wave number can take arbitrary values 
$k>\vartheta$. The S-matrix in an angular momentum channel $j$ is easily read off
from \eqref{WtoM},
\be \label{SMatBos}
S_j(k) \=\frac{\Gamma(\nu+\mu+\frac12)}{\Gamma(\nu-\mu+\frac12)} 
\= \frac{\Gamma\left( j+1- \frac{\I (R^2 k^2-2q^2)}{2R\sqrt{k^2-\vartheta^2}}\right)}
{\Gamma\left( j+ 1+\frac{\I (R^2 k^2-2q^2)}{2R\sqrt{k^2-\vartheta^2}}\right)} \,.
\ee
The density of states in the continuum (relative to the density of states for a free particle in $\IR^3$) is related to the 
phase of the S-matrix via $\rho(k) \de k = \frac{1}{\pi} \de[ \Im \log S(k)]$.
The thermal partition function for a spinless mode, including contributions from the continuum, is then
\be \label{SpecAsymBos} 
\begin{split}
\Tr\, e^{-\beta\cH} \= & \, \Theta(R|\lambda|-|q|)\, \sum_{\ell=0}^{\infty}  \sum_{n=0}^{\infty} (2|q|+2\ell+1)\, 
e^{-\beta R k_n^2/2}  \\ & \, + \sum_{\ell=0}^{\infty} (2|q|+2\ell+1)\, 
\int_{k=\vartheta}^{\infty} \frac{\de k\, \partial_k}{2\pi\I}\left[
\log\frac{\Gamma\left( |q|+\ell+1- \frac{\I (R^2k^2-2q^2)}{2R\sqrt{k^2-\vartheta^2}}\right)}
{\Gamma\left( |q|+\ell+ 1+\frac{\I (R^2 k^2-2q^2)}{2R\sqrt{k^2-\vartheta^2}}\right)}\right]\, e^{-\beta R k^2/2} \,.
\end{split}
\ee
It is worth noting that this expression is {formal} since the sum over the orbital angular
momentum $\ell$ diverges. We shall regulate this divergence by imposing a cut-off at~$\ell\leq \ell_{m}$.

\subsection{Supersymmetric quantum mechanics \label{susyQM}}

Taking into account fermionic zero-modes associated to the supersymmetries broken by the two
dyons, the classical dynamics must be described by  a supersymmetric extension of the previous model  
with 8 supercharges~\cite{Bak:1999da}. One way to find the supersymmetric extension of the 
Lagrangian~\eqref{LagGH} is by dimensional reduction of a two-dimensional $(4,4)$ sigma model on 
a hyperK\"ahler manifold. As shown in \cite{AlvarezGaume:1981hm,AlvarezGaume:1983ab}, such
a model can be deformed by adding a potential proportional to the norm squared of a tri-holomorphic 
vector field. Alternatively, one may start from the undeformed model in two-dimensions
but perform the dimensional reduction with  Scherk-Schwarz twist \cite{Harvey:2014nha}. 
The resulting one-dimensional 
model admits a supersymmetry algebra with a central term \cite{Bak:1999da},
\be
\{Q^{\mu}_{\a} , Q^{\nu}_{\b} \} \= 2 \d^{\mu\nu} \Bigl( \d_{\a\b} \, \CH - \s^{1}_{\a\b} \, \CZ \Bigr) \,,
\ee
where the indices $\alpha,\beta$ run over $\{1,2\}$ while the indices $\mu,\nu$ run over $\{1,\dots, 4\}$, 
corresponding to the four directions on the tangent space of the HK manifold. Defining 
$Q_\pm^\mu=(Q_1^\mu\pm Q_2^\mu)/\sqrt{2}$, this can be rewritten as
\be \label{susyalg}
\{ Q_{\pm}^\mu, Q_{\pm}^\nu \} \= 2 \d^{\mu\nu}\,  ( \cH \mp \cZ ) \, ,\qquad \{Q_+^\mu, Q_-^\nu\} \= 0 \, .
\ee
In view of their two-dimensional origin, we shall refer to $Q_+^\mu$ and $Q_{-}^\mu$
as the right-moving and left-moving supercharges, respectively. In addition to the usual 
fermionic parity $(-1)^F$, which anticommutes with both $Q_+^\mu$ and $Q_{-}^\mu$,
the model admits two $\IZ_2$-gradings\footnote{Upon representing spinors 
on $\MTN$ as multi-forms, the usual fermionic parity is $(-1)^F=(-1)^k$ where $k$ is the form degree,
while $(-1)^{F_+}=\star$ is the Hodge star while $(-1)^{F_-}=\star (-1)^k$.}
 which we shall denote
by $(-1)^{F_\pm}$, such that $(-1)^F=(-1)^{F_+}(-1)^{F_-}$. The operators $(-1)^{F_\pm}$
anticommute with $Q_\pm^\mu$ but commute with $Q_\mp^\mu$, in line with
the fact that they descend from the fermionic parities on the right-moving and 
left moving  side in two-dimensions. 

Energy eigenstates which saturate the BPS bound $E\geq |\cZ|$ 
are either annihilated by $Q^\mu_+$, $\mu=1,\cdots, 4$ (whenever $E=\cZ>0$) or by $Q^\mu_-$, 
$\mu=1,\cdots, 4$ (whenever $E=-\cZ>0$). Unless $|\cZ|=0$, some of the supersymmetries are always broken, 
so the Witten index $\cI=\Tr (-1)^F$ always vanishes. In contrast, the indices $\cI^{\pm}=\Tr(-1)^{F_{\pm}}$ receive 
non-zero contributions  from short multiplets annihilated by~$Q_\pm$~\cite{Stern:2000ie}. 

For the model \eqref{hamTNdef} of interest, the central charge is~$\CZ\= \l q$, 
which we assume to be non-zero. The classical ground states described in \eqref{rcases}
lead to BPS states annihilated by $Q_+$ when $\lambda q>0$, or by $Q_-$ when $\lambda q<0$. 
In either case, they obtain 4 fermionic zero-modes from the broken supersymmetries of the 
quantum mechanics describing the relative motion (as well as another 8 from the center-of-mass motion, 
reproducing the 12 fermionic zero modes of a $\frac14$-BPS bound state in the four-dimensional $\cN=4$
theory). Moreover, the highest weight vector in the supersymmetric multiplet 
carries angular momentum $|q|-\frac12$, with $|q|$ originating from the magnetic term in 
\eqref{Jmag} and $-\frac12$ from the spin degrees of freedom.
It follows that the indices are given by
\be
\label{Indpm}
\cI \= 0 \, ,\qquad \cI^{\pm} \= 4 \left[ |q| \pm q \, \sign(R\lambda\mp q) \right] \, .
\ee
These indices agree with the Dirac indices computed by localization with respect to the action of the 
Killing vector $\partial_\psi$ in \cite{Stern:2000ie}.

One can refine these indices by introducing a fugacity conjugate to conserved charges commuting
with the supercharge as follows.  Using the terminology of the two-dimensional (4,4) sigma model,  
we first note that the algebra \eqref{susyalg} is invariant under independent $SO(4)$ rotations
of the left and right-moving charges. These are {\it a priori} outer automorphisms of the algebra, but it turns out that 
certain combinations are symmetries of the Hamiltonian.
Writing $SO(4)=SU(2)\times \wt{SU(2)}$ on the right-moving side, we define~$J_+$ and~$I_+$
as, respectively, half the sum and half difference of the Cartan generators of~$SU(2)$ and~$\wt{SU(2)}$. 
Similarly,  we define~$J_-$ and~$I_-$ as half the sum and difference of the 
two Cartan generators on the left-moving side. The operators $(-1)^{2J_\pm}$ are the 
$\IZ_2$ gradings mentioned previously, while $J=J_++J_-$ is identified with the Cartan generator of the~$SU(2)$ 
rotational isometry of the Taub-NUT space, corresponding to the 
physical angular momentum of the two-centered system. In addition, there is a conserved charged 
$q$ corresponding to translations along the circle direction $\psi$.

The representations of the supersymmetry algebra \eqref{susyalg} are obtained by tensoring
representations of the left-moving and right-moving algebras. If $E>|\cZ|$, the irreducible
representations on both sides have dimension 4, and carry the charge assignments 
given in  Table~\ref{Rlongrep}. 
Using the fact that $\Tr(-1)^{2J_\pm} y^{2(J_\pm+I_\pm)}=0$ on either of these representations,
it is immediate to see that the resulting long representations, of dimension 16, do not contribute to either of the following
traces,
\bea
\cI^{+}(\l;y,\vv) &\=& \text{Tr}_q \, (-1)^{2J} \, e^{-\b (\CH- q\lambda)} \,  y^{2(J + I_+)}\, e^{4\pi \I \vv I_-} \,,
\label{defIpy}\\
\cI^{-}(\l;y,\vv) &\=& \text{Tr}_q \, (-1)^{2J} \, e^{-\b (\CH+ q\lambda)} \,  y^{2(J + I_-)}\, e^{4\pi \I \vv I_+} \,,
\label{defImy}
\eea
where the trace  is taken over the discrete spectrum in the sector with charge~$q$.
If instead $E=q\lambda>0$, the right-moving representation is one-dimensional, 
and carries $I_+=J_+=0$, while the left-moving representation is the one given in  
Table \ref{Rlongrep}. The resulting short representations, of dimension 4,  do not contribute to $\cI^{-}_q$, but it 
does contribute to $\cI^{+}_q$ with a term proportional to 
$\Tr(-1)^{2J_-} y^{2J_-} e^{4\pi \I \vv I_-}=2\cos(2\pi \vv)-y-y^{-1}$. Similarly,
if $E=-q\lambda>0$, the representation on the left-moving side is one-dimensional, 
and carries $I_-=J_-=0$. The resulting short representations do not contribute to $\cI^{+}$,
but it does contribute to $\cI^{-}$, with a term proportional to $2\cos(2\pi \vv)-y-y^{-1}$. In either case,
the result is independent of $\beta$. Using the fact that the highest weight vector in the
representation carries angular momentum $|q|-\frac12$, we find
\be
\label{Ipres}
\cI^{+}(\l;y,\vv)  \= 
\frac12 \bigl(\sign(q) \, \sign(q-R\lambda)-1 \bigr) \, \left[ \chi_{|q|}(y) - 2\cos(2\pi \vv)\, 
\chi_{|q|-\half} (y)+ \chi_{|q|-1}(y) \right] \, ,
\ee
where $\chi_j(y)=\frac{y^{2j+1}-y^{-2j-1}}{y-1/y}$ is the character
of a spin $j$ representation of $SU(2)$ (we set $\chi_j=0$ whenever $j<0$).  In this expression, the prefactor 
vanishes unless $q(R\lambda-q)>0$, in which case it gives $-1$. Note that this result vanishes at $y=1$, $\vv=0$,
{ in agreement of the vanishing of the Witten index $\cI=0$. 
However its second derivative with respect to $y$
\be 
\label{Ipfp}
\cI^{+}(\l)  \=  -2\left[  \left( y \frac{\de}{\de y} \right)^2 \cI^{+}(\l;y,0) \right]_{y=1}
\ee
happens to agree} with the result for~$\cI_+$ in~\eqref{Indpm}. 
Similarly, the refined index $\cI^{-}$ is given by
\be
\cI^{-}(\l;y,\vv)  \= 
\frac12 \bigl(\sign(q) \, \sign(R\lambda+q) -1 \bigr) \, \left[ \chi_{|q|}(y) - 2\cos(2\pi \vv)\, \chi_{|q|-\half} (y)+ \chi_{|q|-1}(y) \right] \,,
\ee
whose second $y-$derivative at $y=1$, $\vv=0$, happens to agree
 with the result for~$\cI_-$ in~\eqref{Indpm}. This observation suggests that the exotic
 indices $\cI^{\pm}=\Tr (-1)^{F_\pm}$ may be related to more standard indices, where
 states are counted with the physical fermionic parity $(-1)^F = (-1)^{2J}$.

\begin{table}
\centering
\begin{tabular}{ | c | c | c | c| }
\hline
State & $\phantom{-}J_\pm$ & $\phantom{-}I_\pm$ & $(-1)^{2J_\pm}$   \\
\hline
$| \downarrow \downarrow \rangle$ & $-1/2$ & $\phantom{-}0$ & $-1$ \\
$| \downarrow \uparrow \rangle$  &$\phantom{-}0$  &$-1/2$   &  $+1$ \\
$| \uparrow \downarrow \rangle$  & $\phantom{-}0$ & $+1/2$ &  $+1$ \\
$| \uparrow \uparrow \rangle$  & $+1/2$ & $\phantom{-}0$ &  $-1$ \\
\hline
\end{tabular}
\caption{\emph{Long  representation of the chiral superalgebra.} In this table $\pm$ denote the right- and left-moving sectors. 
The two arrows of the state denote the  eigenvalues under the Cartan generators of~$SU(2)$ and~$\wt{SU(2)}$ ($\uparrow$ 
has value $+\half$ and $\downarrow$ has value $-\half$). 
\label{Rlongrep}}
\end{table}

Rather than considering the refined indices $\cI^\pm(\l;y)$, 
which involve a fugacity both for the angular momentum $J$ and R-charge $I_\pm$,
one may consider the helicity partition function 
\be
\label{helpart}
\cI(\l;y) \= \text{Tr}_q \, (-1)^{2J} \, e^{-\b (\CH- |q\lambda|)} \,  y^{2J} 
\ee
with a fugacity $y$ conjugate to the physical angular momentum.
Unlike the refined indices \eqref{defIpy}, \eqref{defImy}, this trace receives contributions from
long representations, given by
\be
\label{heltr}
\sum_{I,J} (-1)^{2J} \chi_{|q|+J-\frac{1}{2}+\ell}(y) \= 
\chi_{|q|+\ell+\frac12}  - 4 \chi_{|q|+\ell} + 6 \chi_{|q|+\ell-\frac12}
-4 \chi_{|q|+\ell-1} + \chi_{|q|+\ell-\frac32}
\ee
where $\ell$ is the orbital angular momentum ({not to be confused with the summation variable
$\ell$ appearing in Section 1}) . Moreover, 
short multiplets contribute in the same way to $\cI(\l;y)$ and $\cI_+(\l;y,0)$ when 
$\lambda q>0$, or to $\cI(\l;y)$ and $\cI^-(\l;y,0)$ when $\lambda q<0$, and in 
both cases carry zero orbital angular momentum. It follows
that the contributions of short multiplets is given by
 \be
 \label{Indy}
 -\frac{ 2 + \sign (q)(\sign(R\lambda-q)-\sign(R\lambda+q))}{2} \, 
\left[ \chi_{|q|}(y) - 2 \chi_{|q|-\frac12}(y) + \chi_{|q|-1}(y) \right] \,,
 \ee
 where the prefactor ensures that $\cI(y)$ vanishes unless $R|\lambda|>|q|$, which is the
 range where bound states exist. 
It is easy to check that
\eqref{heltr} is of order $(y-1)^4$ near $y=1$, while \eqref{Indy} is of order $(y-1)^2$. 
It follows that the second derivative at $y=1$, 
\be
\cI_2 \; \coloneqq \; - \frac12 (y \partial_y)^2 \, \cI(\lambda;y)\vert_{y=1} \,,
\ee
also known as the helicity supertrace, receives only contributions from short multiplets,
coincides with one quarter of the sum of the indices $\cI^{\pm}$ in~\eqref{Indpm},
\be
\label{I2}
\begin{split}
\cI_2
\= 
 2|q| +q\, \sign(R\lambda-q)-q\, \sign(R\lambda+q)  \=
\frac14 \left(  \cI_+ + \cI_-\right) \,.
\end{split}
\ee

As we have shown, the refined indices~$\cI^{\pm}(\l;y,v)$, defined in \eqref{defIpy}, \eqref{defImy} 
as a trace over the discrete spectrum, get contributions only from short BPS states, and are independent 
of the temperature~$\b$. Upon including the contribution of the continuum of scattering states  in the trace, then 
the contribution from bosons and fermions need no longer cancel perfectly, and the resulting indices, 
which we denote by $\wh{\cI}^{\pm}_q(\l;\beta, y,\vv)$, may acquire a dependence on $\beta$.
The density of bosonic and fermionic scattering states can in principle be calculated as in 
Equations~\eqref{SMatBos},~\eqref{SpecAsymBos} from the knowledge of the S-matrix, but this requires 
diagonalizing the action of the Hamiltonian on the 16 helicity states, which is 
cumbersome.\footnote{In Appendix \ref{sec_ham}, we make a tentative guess for the
result of this diagonalisation and compute the resulting helicity supertrace.}
In the  next section, we shall  calculate $\wh{\cI}^{+}_q(\l;\beta, y,\vv)$
using the method of supersymmetric localization. We shall recover the 
contribution of the bound states discussed in this section, as well as  the 
contribution from the continuum, which we compare with the microscopic prediction.

\section{Supersymmetric partition function from localization \label{sec:IndexLoc}}

In this section we compute the refined index~\eqref{defIpy} for the quantum mechanics 
with 8 supercharges described in the previous section, using localization 
in a gauged linear model that flows in the infrared to the model of interest.
We find that the result reproduces the expected contributions of short multiplets in the discrete spectrum, 
plus a $\beta$-dependent contribution which can be ascribed to 
a spectral asymmetry in the continuum. We compare the result with the non-holomorphic
correction term predicted by the microscopic counting. 

\subsection{Localization in the two-dimensional (4,4) sigma model on Taub-NUT}

In the context of two-dimensional (4,4) sigma models, the  elliptic genus of Taub-NUT space $\MTN$
was computed in  \cite{Harvey:2014nha} by localization in a two-dimensional gauged linear model 
which flows to the non-linear (4,4) sigma model on $\MTN$. This gauged linear sigma model simply involves two free
hypermultiplets $(q_1,q_2)\in\IH^2$ and one vector multiplet gauging the non-compact symmetry 
$(q_1,q_2)\to (e^{\I t} q_1, q_2 + \nu t)$ \cite{Tong:2002rq,Harvey:2005ab}. 
At low energy, the model flows to a sigma model on
the hyperK\"ahler quotient $\IH^2///\IR$, which is well-known to be Taub-NUT space. 
In particular, the triholomorphic $U(1)$ isometry and
the rotational $SU(2)$ isometry of $\MTN$ simply descend from the  circle action 
$(q_1,q_2)\to (e^{\I\alpha} q_1,q_2)$ and action of the unit quaternions $(q_1,q_2)\to (p q_1,p q_2 \bar p)$ 
with $p\bar p=1$, which commute with the gauge symmetry \cite[\S 3.1]{Gibbons:1996nt}.
The  authors of \cite{Harvey:2014nha} considered the refined elliptic 
genus\footnote{To match notations, set 
$(Q_f,Q_1,Q_2,Q_R)=(-2q,q_1,-q_2,q_3)$, $z=\vv$, and set $\de u \de \bar u = \de u_1 \de u_2$.}
\be
\label{ECFT}
\cE(\tau;\xi_1,\xi_2,\vv) \; \coloneqq \; \text{Tr}_{\CH_\text{RR}} \, (-1)^{F} \, 
e^{2\pi\I( \tau L_{0} -\bar\tau \wt L_0)} \, 
e^{4\pi\I  \xi_{1} q}\, e^{- 2\pi\I  \xi_{2}(q_{1}+q_{2})} \, e^{- 2\pi\I  \vv q_{3}}\,,
\ee
where $\CH_\text{RR}$ is the Hilbert space on the cylinder in the Ramond-Ramond sector
(including both normalizable states and states in the continuum), 
$L_0, \wt L_0$ are the zero-modes of the Virasoro generators on the cylinder, 
$q$ is the charge under the triholomorphic $U(1)$ action, 
and~$q_1,q_2,q_3$ are the charges under the Cartan generators of
$SU(2)_1 \times SU(2)_2 \times SU(2)_3$, where  $SU(2)_1$ is the action of the unit
quaternions above, while $SU(2)_{2}\times SU(2)_{3}$ is the standard R-symmetry 
of two-dimensional (4,4) sigma models. To see that the observable \eqref{ECFT} is protected,
note that  supercharges transform as $(2,1,2)_-\oplus (2,2,1)_+$ 
under $SU(2)_1\times SU(2)_2 \times SU(2)_3$ (where the subscript indicates the 
two-dimensional helicity), therefore as $(2,2)_-\oplus (1,1)_+ \oplus (3,1)_+$ under 
$SU(2)_L\times SU(2)_3$ where $SU(2)_L$ is the diagonal subgroup of $SU(2)_1\times SU(2)_2$.
Thus, there exists one supercharge which commutes
with $SU(2)_L\times SU(2)_3$, allowing for chemical potentials conjugate to $q_1+q_2$ and 
to $q_3$. Using the localization techniques for $(0,2)$ sigma models developed in \cite{Benini:2013xpa} one finds
\cite[(3.16)]{Harvey:2014nha}:
\be \label{Ecomp}
\cE(\tau;\xi_1,\xi_2,\vv)\=\tR \int_{\cE(\tau)} \frac{\de u_1 \, \de u_2}{\tau_2}\, 
\frac{\theta_1(\tau,u+\xi_1+\vv)\, \theta_1(\tau,u+\xi_1-\vv)}
{\theta_1(\tau,u+\xi_1+\xi_2)\, \theta_1(\tau,u+\xi_1-\xi_2)} \sum_{p,w\in \IZ^2} 
e^{-\frac{\pi \tR}{\tau_2} 
|u+p+\tau w|^2} \,,
\ee
where $u=u_1+\I u_2$, which encodes the holonomies of the vector multiplet,  is integrated over 
the Jacobian torus $ \cE(\tau)=\IC/(\IZ+\tau\IZ)$. The parameter $\tR$, denoted by $g^2$ 
in \cite{Harvey:2014nha}, will be related to the radius $R$ of Taub-NUT shortly.
In this localisation computation, it is important to keep the
parameter $\xi_2$ {non-zero}, since otherwise the two simple poles in the denominator would collide
into a double pole, leading to a logarithmic divergence of the form  $\int \de u \de \bar u \frac{1}{|u|^2}$.
For $\xi_2\neq 0$, the simple poles are integrable, and the result is manifestly holomorphic in $\vv$,
albeit not in $\tau, \xi_1$ nor $\xi_2$.

\subsection{Localization in the quantum mechanics with 8 supercharges on Taub-NUT}
In principle, the localization techniques of \cite{Benini:2013xpa} apply just as well to sigma models 
with 2 supercharges in one dimension \cite{Hori:2014tda}, with several complications due to the fact that
the holonomies of the vector multiplet now live in an infinite cylinder, rather than on a compact torus. 
Alternatively, one may start from the two-dimensional sigma model
and take the limit $\tau_2\to\infty$, so as to remove the contribution of the oscillator 
modes \cite{Hwang:2014uwa,Cordova:2014oxa}. The observable \eqref{ECFT} becomes
\be
\label{ECFTlim}
\wh \cI(\tau_2;\xi_1,\xi_2,\vv) \; \coloneqq \; \text{Tr} \, (-1)^{F} \, e^{-4 \pi \tau_2 \CH} \, 
e^{ 4\pi\I  \xi_{1} q}\, e^{- 2\pi\I  \xi_{2}(q_{1}+q_{2})} \, e^{- 2\pi\I  \vv q_{3}}\, ,
\ee
where $\cH=\frac12(L_0+\wt L_0)$ is the Hamiltonian for the zero-modes. 
Setting $\beta=4\pi\tau_2$,  $y=e^{-2\pi\I\xi_2}$, $\xi_1=\xi_1^r+ \I\lambda\tau_2$, and  identifying 
\be
\label{idenq}
(-1)^{F}\=(-1)^{2J}\ ,\quad q_1\=2J \,, \quad  q_2\=2I_+ \,, \quad q_3\=2 I_-\ ,
\ee
{we recognize the generating function 
\be
\wh \cI(\tau_2;\xi_1,\xi_2,\vv) \=\sum_{2q\in\IZ} e^{4\pi\I  \xi_{1}^r q} \, \wh\cI_q^{+}(\lambda;\beta,
y,\vv)
\ee
of the indices \eqref{defIpy} discussed in the previous section---where the trace in \eqref{ECFTlim} 
 {\it a priori} includes contributions both from normalizable states and from the continuum. }
 The identification $\Im(\xi_1)=\tau_2\lambda$ is motivated by the fact for this choice, 
 the first two exponential factors in \eqref{ECFTlim} recombine into $e^{-\beta (\cH-\cZ)}$ with central 
 charge $\cZ=q\lambda$, as in \eqref{ECFTlim}. The fact that switching on an imaginary part for the chemical 
 potential $\xi_1$ conjugate to the momentum along the triholomorphic isometry induces
a scalar potential proportional to the square of the Killing vector is not obvious and will be justified a posteriori.

Taking the limit $\tau_2\to\infty$ in \eqref{Ecomp},  the Jacobi theta function reduces
to a trigonometric function $\theta_1(\tau,u)\to 2q^{1/8}\sin\pi u$, the contributions of $w\neq 0$ become negligible while  
the integral over the torus $\cE(\tau)$ 
reduces to an integral over  a cylinder of unit radius. We thus arrive at
\be \label{Idef}
\wh \cI(\tau_2;\xi_1,\xi_2,\vv) \= \frac{\tR}{\tau_2} \int_{[0,1] \times \IR} \frac{\de u_1 \, \de u_2}{\tau_2}\,
\frac{\sin\pi(u+\xi_1+\vv) \, \sin\pi(u+\xi_1-\vv)}{\sin\pi(u+\xi_1+\xi_2)\,\sin\pi(u+\xi_1-\xi_2)} \,
\sum_{p\in \IZ} e^{-\frac{\pi \tR |u-p|^2}{\tau_2}} \,,
\ee
where $u_1 \in [0,1]$, $u_2 \in \IR$. As in \eqref{Ecomp}, it is important to keep $\xi_2\neq 0$
in this computation, since otherwise the double pole would lead to a logarithmic divergence.\footnote{{This divergence disappears if one takes both $v=\xi_2=0$, in which case 
the elliptic genus \eqref{Idef} reduces to the Euler number of $\MTN$, which is equal to one.}}
As a result, \eqref{Idef} is manifestly holomorphic in $\vv$ but not in $\xi_1,\xi_2$.

\subsection{Extracting the Fourier coefficients \label{sec_four}}

In order to extract the Fourier coefficients of \eqref{Idef} with respect to $\xi_1$, 
we first do a  Poisson resummation over $p$, obtaining
\be\label{Indpoiss2}
\begin{split}
&\wh \cI(\tau_2;\xi_1,\xi_2,\vv) \= \\
& \qquad  \sqrt{\frac{\tR}{\tau_2}} \int_{[0,1] \times \IR} \de u_1 \, \de u_2\,
\frac{\sin\pi(u+\xi_1+\vv) \, \sin\pi(u+\xi_1-\vv)}{\sin\pi(u+\xi_1+\xi_2)\,\sin\pi(u+\xi_1-\xi_2)} \,
\sum_{2q\in \IZ} e^{-\frac{\pi \tR u_2^2}{\tau_2} -\frac{4 \pi \tau_2 q^2}{\tR} 
- 4\pi\I u_1 q}\,.
\end{split}
\ee
where the dual summation variable is denoted by $q\in\IZ/2$ for later convenience.
Let us now find the Fourier expansion in~$u_1$ for the ratio 
of sine functions in the integrand. The denominator can be written as:
\be
\frac{1}{\sin\pi(u+\xi_1+\xi_2) \, \sin\pi(u+\xi_1-\xi_2) } \= 
\frac{1}{\sin2\pi\xi_2} \biggl( \, \frac{\cos\pi(u+\xi_1+\xi_2)}{\sin\pi(u+\xi_1+\xi_2)} 
\, - \, \frac{\cos\pi(u+\xi_1-\xi_2)}{\sin\pi(u+\xi_1-\xi_2)} \, \biggr) \,.
\ee
Each term can be expanded separately, in an appropriate regime, 
using the formula
\be
\frac{\cos\pi z}{\sin\pi z} \= -\I \,\frac{1+e^{2\pi\I z}}{1-e^{2\pi\I z}} 
\= -\I \,\sum_{n\in\IZ} \, ( \sign(n)+\sign(\text{Im}(z)) )\, e^{2\pi\I n z} \,,\qquad \text{Im}(z) \neq 0 \,,
\ee
with $\sign(0)=0$. The numerator can be written as:
\be
\sin\pi(u+\xi_1+\vv) \sin\pi(u+\xi_1-\vv) \= \frac12 \, \bigl( \cos2\pi \vv - \cos2\pi(u+\xi_1)\bigr) \,.
\ee
Putting these formulae together  we obtain:
\be \label{sinformlua}
\begin{split}
\frac{\sin\pi(u+\xi_1+\vv) \sin\pi(u+\xi_1-\vv)}{\sin\pi(u+\xi_1+\xi_2) \, \sin\pi(u+\xi_1-\xi_2) } & \= 
\bigl(e^{2\pi\I(u+\xi_1)}  -2\cos2\pi \vv + e^{-2\pi\I(u+\xi_1)} \bigr) \; \times  \\
& \sum_{n\in\IZ} \, \biggl[ \, \bigl( \sign(n)+\sign (u_2+\xi_1^i+\xi_2^i)\bigr) \, \frac{e^{2\pi\I n(u+\xi_1+\xi_2)}}{(-2\I) \sin2\pi\xi_2} \,\biggr]_+\,,
\end{split}
\ee
where $\xi_1=\xi_1^r+\I\xi_1^i$, $\xi_2=\xi_2^r+\I \xi_2^i$, and {the notation~$[\,\cdot \,]_+$ 
denotes the even part of a function with respect to~$\xi_2$, namely 
\be
[f(\xi_2) ]_+\coloneqq \frac12[f(\xi_2)  + f(-\xi_2) ]\ .
\ee 
}

We want to rewrite this expression as a Fourier expansion in $\xi_1^r$.
The effect of pulling the three terms in the first parenthesis inside the summation symbol is to shift the value of~$n$
in~$e^{2\pi\I n(u+\xi_1)}$ to~$n+1$, $n$, $n-1$, respectively.  For~$|n|>1$, this shift can be  
absorbed by a corresponding change of the summation variable, because~$\sign(n)=\sign(n\pm1)$ for 
these values. {For the remaining values~$n=0,\pm 1$, this shift changes the 
expression, but by odd function of~$\xi_2$ 
which does not contribute to the even part.} We thus arrive at the expansion:
\be \label{sinformlua2}
\begin{split}
& \frac{\sin\pi(u+\xi_1+\vv) \sin\pi(u+\xi_1-\vv)}{\sin\pi(u+\xi_1+\xi_2) \, \sin\pi(u+\xi_1-\xi_2) } \= \\
&\; \sum_{n\in\IZ} \, e^{2\pi\I n(u+\xi_1)} \, 
 \biggl[ \, \bigl( \sign(n)+\sign (u_2+\xi_1^i+\xi_2^i)\bigr) \, 
 \tfrac{\bigl( e^{2\pi\I (n-1) \xi_2} 
 -2\cos2\pi \vv  \, e^{2\pi\I n \xi_2}  + e^{2\pi\I (n+1) \xi_2}  \bigr)  }{(-2\I)\sin2\pi\xi_2} \,\biggr]_+ \,.
\end{split}
\ee

Now, the integral over $u_1$ in~\eqref{Indpoiss2} identifies the summation variable $n$  
with~$2q$. The integral over $u_2$ splits into two pieces---the first one, 
proportional to~$\sign(n)$ is gaussian, and the second part can be computed using
\be
\sqrt{\frac{\tR}{\tau_2}} \int_\IR \de u\, \sign(u+x)\, e^{-\frac{\pi \tR u^2}{\tau_2}-4\pi q u -\frac{4\pi q^2\tau_2}{\tR}}  \= 
 \erf \Bigl( \sqrt{\frac{\pi \tR}{\tau_2}} x  -\sqrt{\frac{4\pi \tau_2}{\tR}} q \Bigr) \,.
\ee
In this way we arrive at the Fourier expansion of \eqref{Idef} with respect to $\xi_1^r$:
\be  \label{Iresult}
\wh \cI(\tau_2;\xi_1,\xi_2,\vv) \= \sum_{2q \in \IZ} \, \wh \cI_q(\tau_2;\xi_1^i,\xi_2,\vv) \, e^{4\pi \i \,q \,\xi_1} \,,
\ee
with
\be \label{Iqresult}
\begin{split}
&\wh  \cI_q(\tau_2;\xi_1^i,\xi_2,\vv)   \=  \\
& \; 
\biggl[ \Big( \sign(q) - \erf \Bigl(\sqrt{\tfrac{4\pi \tau_2}{\tR}} q - \sqrt{\tfrac{\pi \tR}{\tau_2}} (\xi_1^i+\xi_2^i) \Bigr) \Bigr)
\tfrac{\bigl( e^{2\pi\I (2q-1) \xi_2}  -2\cos2\pi \vv  \, e^{4\pi\I q \xi_2}  + e^{2\pi\I (2q+1) \xi_2}  \bigr)  }{(e^{-2\pi\i \xi_2} - e^{2\pi\i\xi_2} )} 
\biggr]_+ \,.
\end{split}
\ee
The expression~\eqref{Iresult} is then the result for the refined index defined in \eqref{defIpy},
where the trace includes both discrete states and states in the continuum.

\subsection{Interpreting the result}
Performing identifications anticipated above \eqref{idenq}, assuming for the moment 
that $\xi_2$ is real (i.e. $\xi_2^i=0$) and
 further setting $\tR=2R$, the result \eqref{Iqresult} becomes 
\be \label{Iqresult1}
\begin{split} 
\wh  \cI_q(\tau_2,\lambda;\xi_2,\vv)  =
\left[\sign(q) \, \erf \Bigl( \sqrt{\tfrac{\beta}{2R}}(q-R\lambda) \Bigr)-1 \right]\, 
\left[ \chi_{|q|}(y) - 2\cos(2\pi \vv)\, \chi_{|q|-\half} (y)+ \chi_{|q|-1}(y) \right] 
 \end{split}
\ee
with $y=e^{-2\pi\I\xi_2}$ and $\beta=4\pi\tau_2$. 
In the limit $\beta\to+\infty$, this reduces to 
\be \label{Iqresultlim}
\begin{split}
\wh  \cI_q(\lambda;\xi_2,\vv)   \= & \; 
 \left[ \sign\left(q\right)\, \sign\left(q-R\lambda\right)-1 \right]\, 
\left[ \chi_{|q|}(y) - 2\cos(2\pi \vv)\, \chi_{|q|-\half} (y)+ \chi_{|q|-1}(y) \right] \ ,
 \end{split}
\ee
in perfect agreement with the result \eqref{Ipres} for the contributions of short multiplets
in the discrete spectrum (a similar observation was made in  \cite[Equation~(5.15)]{Harvey:2014nha}). 
Interestingly, the error function in \eqref{Iqresult1} also shows up 
with the same argument in the result for the helicity supertrace \eqref{whI2b} computed 
in Appendix \ref{sec_ham}, and it ensures that the result is smooth as a function of $\lambda$,
even at $\lambda=0$ where the potential disappears.
It is also worth noting that \eqref{Iqresultlim} vanishes at $y=1$, however this is only so if this value
is approached along the unit circle $|y|=1$. If we allow $\xi_2$ to have a non-zero imaginary part, 
then the result \eqref{Iqresult} is in fact divergent at $\xi_2=0$, reflecting the logarithmic divergence
of the integral \eqref{Idef} at that value. In fact, just as the imaginary part of $\xi_1$ is related to the
coefficient $\lambda$ of the scalar potential on Taub-NUT, one might expect that a non-zero value
of $\xi_2^i$ may have a similar effect of inducing a scalar potential, and change the classical dynamics 
of the system.

Let us now extract the index $\wh \cI_q^+$ by taking two derivatives with respect to $\xi_2$ 
before setting $\xi_2=0$ as in \eqref{Ipfp}, i.e.
\be 
\label{Iplusqresult0}
\wh \cI_q^+(\tau_2;\lambda) \; \coloneqq \;  \frac{1}{2\pi^2} \frac{\de^2}{\de\xi_2^2} \, \wh \cI_q(\tau_2;\lambda,\xi_2,\vv) \Big|_{\xi_2=\vv=0} \,.
\ee
If we restrict $\xi_2$ to lie along the imaginary axis ($\xi_2=\I \xi_2^i$), we find 
\be 
\label{Iplusqresult}
\begin{split}
\wh \cI_q^+(\tau_2;\lambda) \=  4\,|q| - 4\,q\,  \erf\,\biggl( \sqrt{\frac{2 \pi \t_2}{R}} (q-\lambda R)  \biggr) 
- \frac{2}{\pi}\sqrt{\frac{2R}{\t_2}} e^{-2 \pi \t_2(q-\lambda R)^2/R} \,.
\end{split}
\ee
This is precisely the function~$-4\, \wh{a}_{\ell}^\text{attr}(\tau_2,\wt u_2)$ in Equation~\eqref{amattr},
upon identifying $m=2R$, $\wt u_2 = -m \lambda$ and~$\ell=2q$.
The overall factor of $-4$ is due to our choice of normalization, which was tailored to match
the indices $\cI^{\pm}$ in \eqref{Indpm} in the limit where $\tau_2\to\infty$.
We note that other ways of treating the derivative $\frac{\de}{\de\xi_2}$ in~\eqref{Iplusqresult0} 
would give a different coefficient for the Gaussian term in~\eqref{Iplusqresult}.
At the moment we do not have a physical justification for the prescription used above, which 
seems to be required for modularity.

\subsection{Supersymmetric quantum mechanics with four supercharges \label{sec_loc4}}
Here we briefly discuss the index in the supersymmetric quantum mechanics obtained by 
reducing the (0,4) sigma model on Taub-NUT space, which provides an alternative
description of the quantum mechanics of two BPS black holes in $\cN=2$ string vacua. 
The elliptic genus in this model was
computed using the same localization techniques in \cite[(6.11)]{Harvey:2014nha}.
Including the contribution of the left-moving fermions, we arrive at 
\be
\label{Ell04}
\cE'(\tau;\xi_1,\xi_2)\=\tR \int_{\cE(\tau)} \frac{\de u_1 \de u_2}{\tau_2}\, 
\frac{\eta^6}
{\theta_1(\tau,u+\xi_1+\xi_2)\, \theta_1(\tau,u+\xi_1-\xi_2)} \sum_{p,w\in \IZ} e^{-\frac{\pi \tR}{\tau_2} 
|u+p+\tau w|^2} \,,
\ee
where $\xi_1$ couples to the $U(1)$ charge conjugate to the tri-holomorphic
isometry, and $\xi_2$ couples to a linear combination of Cartan generators for the 
rotational isometry and R-symmetry. As before, $\xi_2$ must be kept non-zero
in order for the integral to be well-defined. 
In the limit $\tau_2\to\infty$, this becomes
\be
\label{Ind2}
\wh \cI' (\tau_2;\xi_1,\xi_2)\=\frac{\tR}{\tau_2} \int_{[0,1] \times \IR} \de u_1 \de u_2\,
\frac{\sum_{p\in \IZ} e^{-\frac{\pi\tR |u+p|^2}{\tau_2}} }{4\sin(u+\xi_1+\xi_2)\,\sin(u+\xi_1-\xi_2)}\,.
\ee
The Fourier expansion with respect to $\xi_1^r$ can be computed using the same methods
as in~\S\ref{sec_four}. Upon identifying $\xi_1^i=\tau_2 \lambda$ as before, and taking
the limit~$\xi_2\to 0$ keeping $\xi_2$ purely imaginary we find
\be 
\label{Iplusqresult4}
\begin{split}
\wh \cI'_q(\tau_2;\lambda) \=  |q| - q\,  \erf\,\biggl( \sqrt{\frac{2 \pi \t_2}{R}} (q-\lambda R)  \biggr) 
- \frac{1}{2\pi}\sqrt{\frac{2R}{\t_2}}
e^{-2 \pi \t_2(q-\lambda R)^2/R} \,\ ,
\end{split}
\ee
i.e. precisely the same result~\eqref{Iplusqresult}
as in the model with 8 supercharges, up to an overall factor of 1/4. In particular, in contrast to the
model studied in \cite{Pioline:2015wza}, the contribution from the continuum produces both
a term proportional to the complementary error function, as well as a Gaussian term,
which is in fact necessary for the modular invariance of the generating function
of MSW invariants \cite{Manschot:2009ia,Alexandrov:2016tnf}.

\section{Discussion\label{sec_disc}}

In this paper we studied the supersymmetric quantum mechanics of a particle moving in Taub-NUT space $\MTN$, as a model for the relative dynamics of two-black-hole bound states in~$\CN=4$ string theory. 
We analyzed this system both from a Hamiltonian  viewpoint and by using localizing the
functional integral. 
The spectrum of the theory consists of a discrete part, corresponding to bound states, as well as a 
continuum part, corresponding to scattering states. Our main goal was to compare the
contribution of the continuum  with the non-holomorphic completion required 
for modularity of the generating function of black hole degeneracies in the microscopic analysis.

We mainly focussed on the supersymmetric index~$\wh \cI(\tau_2;\xi_1,\xi_2,\vv)$ where the parameter~$\t_2$ 
couples to the Hamiltonian, $\xi_1$ couples to the~$U(1)$ charge $q$ under the triholomorphic isometry 
of  $\MTN$, $\xi_2$ to a combination of the Cartan generator of the~$SU(2)$ rotational isometry 
and an R-charge $q_2$, and~$\vv$ to different R-charge $q_3$ in the supersymmetric quantum
mechanics. The imaginary part of $\xi_1$ is proportional to the coefficient $\lambda$ of the scalar potential 
which deforms the geodesic motion on $\MTN$, while preserving all supersymmetries. 
Using the Hamiltonian formulation of the model, we computed the contribution of the discrete states to 
the above refined index, as well as to other indices and helicity supertraces. 
We recovered the same result using supersymmetric localization in the  functional integral,
along with contributions from the continuum of scattering states. 
The main result is summarized in Equations~\eqref{Iresult}, \eqref{Iqresult}.
The discrete part of this result agrees with the Hamiltonian computation 
upon identifying ~$\Im(\xi_1)=\t_2 \lambda$.

Upon computing the second Taylor coefficient in~$\xi_2$ at~$\vv=0$, assuming  the chemical potential~$\xi_2$ 
to be purely imaginary, we found that $\wh \cI(\tau_2;\xi_1,\xi_2,\vv)$ precisely
reproduces the Fourier coefficient~$\wh{a}_{\ell}^\text{attr}(\tau_2,\wt u_2)$ in \eqref{amattr}
appearing in the modular completion of the generating function \eqref{ALhat13var} of the microscopic 
degeneracies---a generalization of the usual generating function  \eqref{defGsmhat}  involving two elliptic parameters $z, \wt z$.
The parameter $\wt u_2=\Im(\wt z)/\tau_2$ on the microscopic side is identified with $\lambda$,
whereas the parameter $u_2=\Im(z)/\tau_2$ must be taken in the attractor chamber in order
to match the quantum mechanics result. The function~$\wh{a}_{\ell}^\text{attr}(\tau_2,\wt u_2)$ 
encodes the modular completion of the original one-parameter generating function  $\ALhat(\t,z)$,
in a subtle manner which combines the limits $\wt u_2 \to 0$ and $|\wt u_2|\to\infty$ as discussed
at the end of \S\ref{sec_AL3}.

Our analysis raises several puzzles and open questions. 
First, it would be interesting to have an independent computation of the continuum contribution
to the refined index using Hamiltonian methods. In an appendix, we outline such a computation
for the helicity supertrace, but it remains to extend this approach to the case of the refined index.
Second, it would be useful to justify why the imaginary part of the chemical potential $\xi_1$ 
induces a scalar potential on Taub-NUT space, and whether the imaginary part of $\xi_2$
has a similar effect. Third, we have observed certain relations between the indices 
$\cI^{\pm}=\Tr (-1)^{F_\pm}$, the helicity supertrace $\cI_2$ and the second derivatives 
of $\cI^{\pm}(y,v)$ with respect at $y=1, v=0$ at the level of the discrete state contributions,
and it would be interesting to establish if these relations continue to hold beyond the limit
$\beta\to\infty$. 

As for the comparison with the generating function of microscopic degeneracies of $\cN=4$ dyon bound states, 
it is satisfying that the quantum mechanics produces the correct
non-holomorphic completion term of the full three-variable Appell-Lerch sum
\eqref{ALhat13var}, but it is puzzling that it matches the  bound
state contributions only in the attractor chamber $u_2=-\ell/2m$ (albeit for all values of $\wt u_2$).
This is presumably due to the fact that we have not found a natural r\^ole for the chemical
potential $u_2=\Im(z)/\tau_2$ in the quantum mechanics. It would be interesting
to understand the physical relevance of the three-parameter generating function defined
in~\eqref{Aprm}, and see whether a similar refinement exists for the generating
function of single-centered $\cN=4$ black holes.
Another issue worth clarifying is the dependence of the result~\eqref{Iplusqresult}
on the direction of the derivative in Equation~\eqref{Iplusqresult0}.

Finally, it is interesting to note that the quantum mechanics on Taub-NUT with 4 supercharges 
provides an alternative description of the dynamics of two-centered black holes in $\cN=2$ string
vacua, which is different from the one studied in \cite{Denef:2002ru},\cite{Lee:2011ph},\cite{Pioline:2015wza}. 
In Section~\ref{sec_loc4} we computed the index using localization, and found that the
result \eqref{Iplusqresult4} contains both a term proportional to the complementary error function, 
also present in~\cite{Pioline:2015wza}, as well as a Gaussian term,
which is in fact necessary for the modular covariance of the generating function
of MSW invariants \cite{Manschot:2009ia,Alexandrov:2016tnf}. It would
be interesting to apply similar localization techniques to the case of multi-centered
black holes, where mock modular forms of higher depth are expected 
to occur \cite{AlexandrovBPtoappear}. Interestingly, such modular objects arise
in the computation of elliptic genera of squashed  toric manifolds \cite{Gupta:2018krl}, and presumably
also in the context of higher rank monopole moduli spaces, which may provide a useful model
for the dynamics of multi-centered black holes.

\section*{Acknowledgements}
We are grateful to Guillaume Bossard, Atish Dabholkar, Rajesh Gupta, Sungjay Lee, 
Jan Manschot, Greg Moore, Caner Nazaroglu, and Piljin Yi for useful discussions over the course
of this project. The research of S.~M.~is supported
by the ERC Consolidator Grant N.~681908, ``Quantum black holes: A
microscopic window into the microstructure of gravity'', and by the STFC grant ST/P000258/1. 
The research of B.~P.~is supported in part by French state funds 
managed by the Agence Nationale de la Recherche (ANR) in the context of the LABEX ILP (ANR-11-IDEX-0004-02, ANR-10- LABX-63).


\appendix

\section{Spectral asymmetry and helicity partition function \label{sec_ham}}

In this section, we compute the helicity partition function
\be
\label{helpart2}
\wh \cI(\l;\beta,y) = \text{Tr} \, (-1)^{2J} \, e^{-\b (\CH- |q\lambda|)} \,  y^{2J} 
\ee
using Hamiltonian methods. This function is not a protected quantity, since $J$ does not commute 
with any supercharge. Moreover, the explicit computation in \eqref{heltr} shows that long multiplets in the discrete spectrum contribute. In contrast, the second derivative at $y=1$,
 \be
 \wh \cI_2(\beta) \; \coloneqq \;  \frac12 (y \partial_y)^2  \, \wh \cI (\lambda;\beta,y)\vert_{y=1}
\ee
receives only contributions from short multiplets. In the zero temperature limit $\beta\to\infty$, $\wh \cI_2(\beta)$
reduces to the helicity supertrace $\cI_2$ in \eqref{I2}, but it may also receive contributions from the continuum 
of scattering states due to a possible  asymmetry between the bosonic and fermionic densities of states. 
These contributions are in fact  necessary in order to ensure that $\wh \cI(\beta)$ is a smooth function 
of the parameter $\lambda$ at finite $\beta$, as required by the Fredholm property of the Hamiltonian $\cH$. 
To compute the spectral asymmetry, we shall follow the same approach as in \cite{Pioline:2015wza}, with a shortcut
to eschew a full analysis of the supersymmetric quantum mechanics. Unfortunately, applying
the same shortcut to the computation of the refined indices introduced
in \eqref{defIpy}, \eqref{defImy} does not seem to give a sensible result, 
so the results
in this appendix should be viewed as heuristic.

In the case with four supercharges, relevant for dyon dynamics 
in $\cN=2$ gauge theories, the  wave-function is a 4-component vector which decomposes under the 
rotation group $SU(2)$ as 2 scalars and one doublet. This model is similar to the one studied
in \cite{Pioline:2015wza}, in fact for $\lambda=0$ it agrees with it  upon rescaling the metric 
on $\IR^4$ by the harmonic function $H$. In addition to the bosonic part  \eqref{hamTNdef},  the Hamiltonian 
also includes couplings between the spin and the magnetic field $\vec B=q \vec r/r^3$ sitting at the origin. 
After decomposing each mode  into a radial and angular part using
spin-weighted monopole harmonics and diagonalizing the resulting radial Hamiltonian, one finds
that the energy levels and density of states in the continuum 
for a mode of helicity $h\in\{0,0,\pm \frac12\}$ are obtained from those  of the bosonic model by replacing
the relation $j=|q|+\ell$, $\nu=j+\frac12$ in \eqref{reljq}, \eqref{munu} by 
\be
j = |q|+h+\ell\ ,\qquad  \nu=j+h+\frac12 \,,
\ee
where $\ell\in\IN$ is still the orbital angular momentum.  In particular, the 
ground state is now obtained by setting $n=\ell=0, h=-\frac12$, and saturates the bound 
$E\geq |q\lambda|$. It now transforms as a representation of 
spin $|q|-\frac12$  under $SU(2)$ , 
and is annihilated by half of the 4 supersymmetries whenever $q\lambda>0$.

In  supersymmetric quantum mechanics with eight supercharges, the wave-function becomes a 
$2^4$-component vector, obtained by tensoring the previous one with a basic 4-dimensional multiplet 
comprising two scalar modes and one spin 1/2. Thus, it  decomposes under the rotation group $SU(2)$ as 5 scalars, 
4 doublets and one triplet, corresponding to helicities $\{h_i, i=1,\dots ,16\} = \{6[0], 4[\pm \frac12], \pm 1\}$ which 
are apparent in the contribution \eqref{heltr} of the long multiplets. In principle, one should again decompose each 
mode into a radial and angular part, and diagonalize the resulting radial Hamiltonian. Rather than carrying out this 
cumbersome procedure, we shall assume that the resulting energy levels and density of states are still obtained 
from those  of the bosonic model by replacing the relation $j=|q|+\ell$, $\nu=j+\frac12$ in \eqref{reljq}, \eqref{munu} by 
\be
j = |q|+h_i+\ell\ ,\qquad  \nu=j+h'_i+\frac12 \,,
\ee
where however $h'_i$ need no longer be equal to $h_i$, due to possible mixing among the various modes of helicity $h$. 
Moreover, we shall assume that for any $i$, $h'_i\in\{0,0,\pm\frac12\}$ is the helicity of the mode of the model with 4 supercharges, 
which leads to the mode of helicity $h_i$ after tensoring with the basic multiplet. This ensures that the BPS ground state, obtained 
by tensoring the ground state $n=\ell=0,h'=-\frac12$ of spin $|q|$ by the basic multiplet,  now includes one multiplet of 
spin $|q|-\frac12$, 2 multiplets of spin $|q|$ and one multiplet of spin $|q|+\frac12$.

Identifying the fermionc parity $(-1)^F$ with $(-1)^{2h}$, we obtain, for the model with 8 supercharges,
\be
\label{Ibetay}
\begin{split}
\wh\cI(\beta,y) & \= \Tr (-1)^{2J_3}  y^{2J_3} e^{-\beta(\cH-|q\lambda|)}  \\
& \=  \frac12 \left( 2 + \sign (q) \bigl(\sign(R\lambda-q)-\sign(R\lambda+q) \bigr) \right)\ 
\left[ \chi_{|q|} - 2 \chi_{|q|-\frac12} + \chi_{|q|-1} \right] 
\\
&\qquad\quad +\sum_{i=1}^{16} (-1)^{2h_i}  \sum_{\ell=0}^{\infty}
\chi_{|q|+h_i+\ell}\, 
\int_{\vartheta}^{\infty} \de k\, e^{-\frac{\beta R k^2}{2}+\beta |\lambda q|}
\\
&\qquad\qquad \qquad \times \ 
\frac{\partial_k}{2\pi\I}\left[
\log\frac{\Gamma\left( |q|+\ell+2h'_i+1- \frac{\I(R^2 k^2-2q^2)}{2R\sqrt{k^2-\vartheta^2}}\right)}{\Gamma\left( |q|+\ell+2h'_i
+ 1+\frac{\I(R^2 k^2-2q^2)}{2R\sqrt{k^2-\vartheta^2}}\right)}
\right] \,.
\end{split}
\ee
Under our assumptions, the contribution of the continuum in the second line, which we denote 
by $\cJ(\beta,y)$, becomes
\be
\label{Jguess}
\begin{split}
\cJ(\beta,y) \=& \left[\chi_{\frac12}-2\chi_0 \right]  \sum_{\ell=0}^{\infty} \int_{\vartheta}^{\infty} \de k\,
e^{-\frac{\beta R k^2}{2}+\beta |\lambda q|}
\\&
\times \frac{\partial_k}{2\pi\I}\left[
\chi_{|q|+\ell-\frac12} \log \Gamma\left(\tfrac{z_\ell}{\bar z_{\ell}}\right)
-2\chi_{|q|+\ell} \log \Gamma\left(\tfrac{z_\ell+1}{\bar z_{\ell}+1}\right)
+ \chi_{|q|+\ell+\frac12} \log \Gamma\left(\tfrac{z_\ell+2}{\bar z_{\ell}+2}\right)
\right] \,,
\end{split}
\ee
where we denoted $z_\ell=|q|+\ell- \frac{\I(R^2 k^2-2q^2)}{2R\sqrt{k^2-\vartheta^2}}$.
By construction, the function~$\cJ(\beta,y)$ vanishes at $y=1$ due to the double zero in the 
prefactor~$(\chi_{\frac12}-2\chi_0)$. The second derivative at $y=1$ removes that zero and leaves 
the expression inside the integral evaluated at~$y=1$. In this limit the characters $\chi_j(y)$ 
reduce to $2j+1$,
and using the identity $x\Gamma(x)=\Gamma(x+1)$, 
the three terms in the parenthesis in the second line above add up to:
\be
 -2(|q|+\ell) \log (z_\ell/\overline z_\ell)  + 2(|q|+\ell+1) \log (z_{\ell+1}/\overline z_{\ell+1})  \ .
\ee
Thus, as in~\cite{Pioline:2015wza}, all terms with $\ell>0$ cancel in the second derivative at $y=1$, 
leaving only the~$\ell=0$ term~$\cJ^{(0)}(y,\beta)$. 
Taking two derivatives in $y$ before setting $y=1$, we arrive at
\be
\begin{split}
\label{defJ2}
\cJ^{(0)}_2(\beta) \= &\, \frac12\partial_y^2 \cJ^{(0)}(y,\beta)\bigg{|}_{y=1}  \=
-2|q|\, \int_{k=\vartheta}^{\infty}  
\frac{\de k \, \partial_k}{2\pi\I}\left[
\log\frac{ |q|- \frac{\I(R^2 k^2-2q^2)}{2R\sqrt{k^2-\vartheta^2}}}{ |q|+\frac{\I (R^2 k^2-2q^2)}{2R\sqrt{k^2-\vartheta^2}}}
\right]\, e^{-\frac{\beta R k^2}{2}+\beta |\lambda q|} \\
\=& \, \frac{4 q^2 R}{\pi} \int_{k=\vartheta}^{\infty}  \frac{k(k^2-2\lambda^2)\de k}
{(R^2 k^4-4 q^2 \lambda^2 )\sqrt{k^2-\vartheta^2}} e^{-\frac{\beta R  k^2}{2}+\beta |\lambda q|}  \,.
\end{split}
\ee
The pole in the integrand at $Rk^2=2|q\lambda|$  (which  does not belong to the integration
range) reflects the existence of the BPS bound state with energy $|q\lambda|$. The integral converges both at the lower and upper bounds, but is discontinuous when $\lambda$ crosses the values $\pm |q|/R$, cancelling the discontinuity in the bound state contribution. To expose this discontinuity, it is expedient to decompose the rational function in the integrand according to 
\be
\label{partialfrac}
\frac{1}{R^2 k^4-4 q^2 \lambda^2} \= \frac{1}{4q\lambda} \left[ 
\frac{1}{Rk^2-2 q \lambda} - \frac{1}{Rk^2+ 2 q \lambda} \right] \,.
\ee
Further changing variable to $v=\sqrt{\frac{k^2}{\vartheta^2}-1}$, we find 
\be
\cJ^{(0)}_2(\beta)=\frac14 \bigl(\cJ_+(\beta)+\cJ_-(\beta)\bigr) \,,
\ee
with 
\be
\label{Jintvpm}
\begin{split}
\cJ_{\pm}(\beta) \= & \pm 4 \frac{q|\vartheta|}{\pi \lambda} \int_0^{\infty} \de v 
\frac{R^2\vartheta^2(v^2-1)+2q^2}{R^2\vartheta^2 v^2 + (R\lambda\mp q)^2}
 e^{-\frac{ \beta R\vartheta^2}{2}
(1+v^2)+\beta |\lambda q|} 
\\
\= & \pm 4 \frac{q|\vartheta|}{\pi \lambda} \int_0^{\infty} \de v 
\left[ 1 - \frac{2R\lambda(R\lambda\mp q)}{R^2 \vartheta^2 v^2 + (R\lambda\mp q)^2} \right]
 e^{-\frac{ \beta R\vartheta^2}{2}
(1+v^2)+\beta |\lambda q|} \,.
\end{split}
\ee
The first term in the square bracket gives a Gaussian integral, while the second term can be computed by 
rescaling $v\to v |R\lambda \mp q|/R|\vartheta|$ and using
\be
\label{M1intv}
\int_0^{\infty} \frac{e^{-\pi u^2(v^2+1)}}{v^2+1}  \, \de v
\= \frac{\pi}{2} \, {\rm erfc}(|u|\sqrt{\pi})  \,.
\ee
As a result, we find 
\be
\label{Jpmfin}
\cJ_{\pm}(\beta) \= \mp \,
 4 q\,  
 \left[ \sign(R\lambda\mp q)\, \erfc\left( |R\lambda\mp q| \sqrt{\frac{\beta}{2R}}\right)
- \frac{e^{-\frac{ \beta   (R\lambda\mp q)^2}{2R}}}{\lambda\sqrt{2\pi \beta R }}\,  
 \right]\, e^{\mp \beta \lambda q +\beta |\lambda q|} \,.
\ee
Note that while the Gaussian term is singular at $\lambda=0$, it cancels in the sum $\cJ_++\cJ_-$,
and could in fact be removed by adjusting the constant term in the decomposition \eqref{partialfrac}. 

Reinstating the contribution from the discrete states on the second line of \eqref{Ibetay}, which we can rewrite as 
\be
\cI_2 \=   2 |q| + q\,  \sign(R\lambda-q)  \, e^{\beta( |\lambda q|-\lambda q)}
 \,-\, q \, \sign(R\lambda+q) \, e^{\beta( |\lambda q|+\lambda q)} \,,
\ee
since both exponentials reduce to one when the respective prefactor is non-zero, 
we finally arrive at 
\be
\label{whI2b}
\wh \cI_2(\beta) \= 2 |q| 
+ q\, \erf\left( (R\lambda- q) \sqrt{\frac{\beta}{2R}}\right)\,  e^{\beta( |\lambda q|-\lambda q)}
- q\, \erf\left( (R\lambda+ q) \sqrt{\frac{\beta}{2R}}\right)\,  e^{\beta( |\lambda q|+\lambda q)} \,.
\ee
In the limit $\beta\to\infty$, this reduces to the helicity supertrace \eqref{I2}, 
but is a continuous function of $\lambda$ for any finite value of $\beta$
(albeit not differentiable at $\lambda=0$.) 

It is tempting to identify the contributions $\cJ_{\pm}(\beta)$ with the indices
$\wh \cI^{\pm}(y,\beta)$ at $y=1$. However, they differ from the localisation result 
\eqref{Iplusqresult}, and there is no reason a priori to expect that the helicity supertrace
$\cI(\beta)$ should be related to the sum of the indices $\cI^{\pm}(y,\beta)$,
even though this appears to be the case for the contribution of the discrete spectrum.

%
%

\providecommand{\href}[2]{#2}\begingroup\raggedright\endgroup

\end{document}